\documentclass[twocolumn,secnumarabic,amssymb,showpacs,superscriptaddress, aps, prl]{revtex4-1}

\usepackage{amsmath} 
\usepackage{amsthm} 
\usepackage{amssymb}	
\usepackage{graphicx} 
\usepackage[table]{xcolor}
\usepackage{comment}
\usepackage[export]{adjustbox}
\usepackage{hyperref}
\hypersetup{
	colorlinks  = true,
	citecolor    = blue
}
\usepackage{blindtext}
\usepackage{enumitem}
\usepackage{lipsum}
\usepackage{verbatim}
\usepackage[normalem]{ulem}

\newcommand{\ket}[1]{\left| #1 \right\rangle} 
\newcommand{\Z}{\mathbb{Z}}

\newcommand{\ketbra}[2]{\left| #1 \rangle\langle #2 \right|}

\begin{document}

\title{Probing quantum thermalization of a disordered dipolar spin ensemble with discrete time-crystalline order}
\affiliation{Department of Physics, Harvard University, Cambridge, Massachusetts 02138, USA}
\affiliation{School of Engineering and Applied Sciences, Harvard University, Cambridge, Massachusetts 02138, USA}
\affiliation{Research Centre for Knowledge Communities, University of Tsukuba, Tsukuba, Ibaraki 305-8550, Japan}
\affiliation{Institut f{\"u}r Quantenoptik, Universit{\"a}t Ulm, 89081 Ulm, Germany}
\affiliation{Takasaki Advanced Radiation Research Institute, National Institutes for Quantum and Radiological Science and Technology, 1233 Watanuki, Takasaki, Gunma 370-1292, Japan}
\affiliation{Sumitomo Electric Industries Ltd., Itami, Hyougo, 664-0016, Japan}
\affiliation{Department of Theoretical Physics, University of Geneva, 1211 Geneva, Switzerland}

\author{Joonhee Choi$^{1,2}$}
\thanks{These authors contributed equally to this work} 
\author{Hengyun Zhou$^{1}$}
\thanks{These authors contributed equally to this work} 
\author{Soonwon Choi$^{1}$} 
\author{Renate Landig$^{1}$}
\author{Wen Wei Ho$^{1}$}
\author{Junichi Isoya$^{3}$}
\author{Fedor Jelezko$^{4}$}
\author{Shinobu Onoda$^{5}$}
\author{Hitoshi Sumiya$^{6}$}
\author{Dmitry A. Abanin$^{7}$}
\author{Mikhail D. Lukin$^{1}$}
\email{lukin@physics.harvard.edu}

\begin{abstract}
We investigate thermalization dynamics of a driven dipolar many-body quantum system through the stability of discrete time crystalline order. Using periodic driving of electronic spin impurities in diamond, we realize different types of interactions between spins and demonstrate experimentally that the interplay of disorder, driving and interactions leads to several qualitatively distinct regimes of thermalization. For short driving periods, the observed dynamics are well described by an effective Hamiltonian which sensitively depends on interaction details. For long driving periods, the system becomes susceptible to energy exchange with the driving field and eventually enters a universal thermalizing regime, where the dynamics can be described by interaction-induced dephasing of individual spins. Our analysis reveals important differences between thermalization of long-range Ising and other dipolar spin models.
\end{abstract}

\maketitle
Thermalization is a universal feature of most many-body systems~\cite{deutsch1991quantum, srednicki1994chaos, rigol2008thermalization, d2016quantum, langen2015experimental, kaufman2016quantum}, underlying the applicability of equilibrium statistical mechanics. At the same time, it represents an important limitation for the coherent manipulation of large scale quantum systems in quantum information processing. For these reasons, a detailed understanding of thermalization processes in closed, interacting quantum many-body systems is of great interest to both fundamental and applied science.

Recently, it was demonstrated experimentally that thermalization processes in many-body systems can be significantly slowed, or even halted due to strong disorder \cite{anderson1958absence,gornyi2005interacting,basko2006metal, nandkishore2015many,abanin2018ergodicity,schreiber2015observation, monroe2016MBL,choi2016exploring,kucsko2016critical,roushan2017spectroscopic,wei2018exploring}. The suppression of thermalization allows for novel nonequilibrium states of matter that would otherwise be forbidden in equilibrium. One remarkable example is the discrete time crystal phase in periodically-driven (Floquet) systems \cite{vedika2016phase,else2016floquet,yao2017discrete,von2016absolute,else2017prethermal,ho2017critical}. This phase is
characterized by a spontaneous breaking of the discrete time-translational symmetry of the drive, which is manifested in local observables exhibiting long-lived, robust oscillations at a subharmonic of the fundamental driving frequency. Signatures of discrete time-crystalline (DTC) order have been observed in various experimental platforms such as trapped ions, electronic and nuclear spin ensembles~\cite{zhang2017observation,choi2017observation,rovny2018observation,pal2018observation}. Since the stability of DTC order is closely related to the suppression of thermalization processes, these observations also raise the intriguing possibility of using the DTC signal as a tool to study thermalization dynamics in an interacting many-body system. 

In this Letter, we demonstrate that the stability of DTC order can be used as a sensitive, quantitative probe of thermalization behavior in a quantum many-body dipolar system. Specifically, we coherently manipulate a disordered ensemble of dipolar-interacting spins to engineer Floquet dynamics with three different types of interactions. In all cases, robust, long-lived signatures of DTC order can be observed over some range of parameters. By tuning both the Floquet period and the strength of perturbations, we monitor the corresponding changes in the decay of DTC order that ensue, which allows us to study thermalization dynamics in these systems.

Our experimental observations reveal the presence of three distinct thermalization regimes. In the case where the driving period is short compared to the inverse of disorder strength, DTC order is robust over a wide range of perturbation strengths, and we find that spin  dynamics is well described by an effective, static Hamiltonian which sensitively depends on the details of interactions~\cite{else2017prethermal,abanin2015exponentially,mori2016rigorous,kuwahara2016floquet,abanin2017effective}. Thermalization occurs only via rare resonances that are strongly suppressed by the large disorder \cite{levitov1990delocalization,burin2006energy,yao2014many}. At longer periods, the effective Hamiltonian description breaks down as the system can exchange energy with the periodic drive, but long-lived DTC order can still persist. This stability can be attributed to critically slow thermalization dynamics arising from the delicate interplay of the long-range nature of interactions with disorder and driving, in agreement with previous observations of a critical DTC regime~\cite{kucsko2016critical,ho2017critical,choi2017observation}. At sufficiently long drive periods, DTC order becomes unstable as the system enters a third thermalizing regime, characterized by universal dynamics independent of the interaction details. This regime can be effectively modeled as individual spins undergoing Markovian dephasing, suggesting that the many-body system serves as its own Markovian bath. However, we find that the crossover to this regime depends strongly on the nature of interactions, indicating differences in thermalization dynamics of long-range Ising and other dipolar spin models~\cite{safavi2018quantum}. Our results have important implications for the dynamical engineering of Hamiltonians~\cite{waugh1968approach,choi2017dynamical}, novel Floquet phases in many-body systems ~\cite{potter2016classification, von2016phase, von2016phase2, PhysRevB.94.125105, else2016classification, nathan2017stability}, with applications to quantum metrology~\cite{choi2017quantum} and quantum simulations~\cite{lanyon2011universal}.

\begin{figure}[t]
\includegraphics[width=.48\textwidth,left]{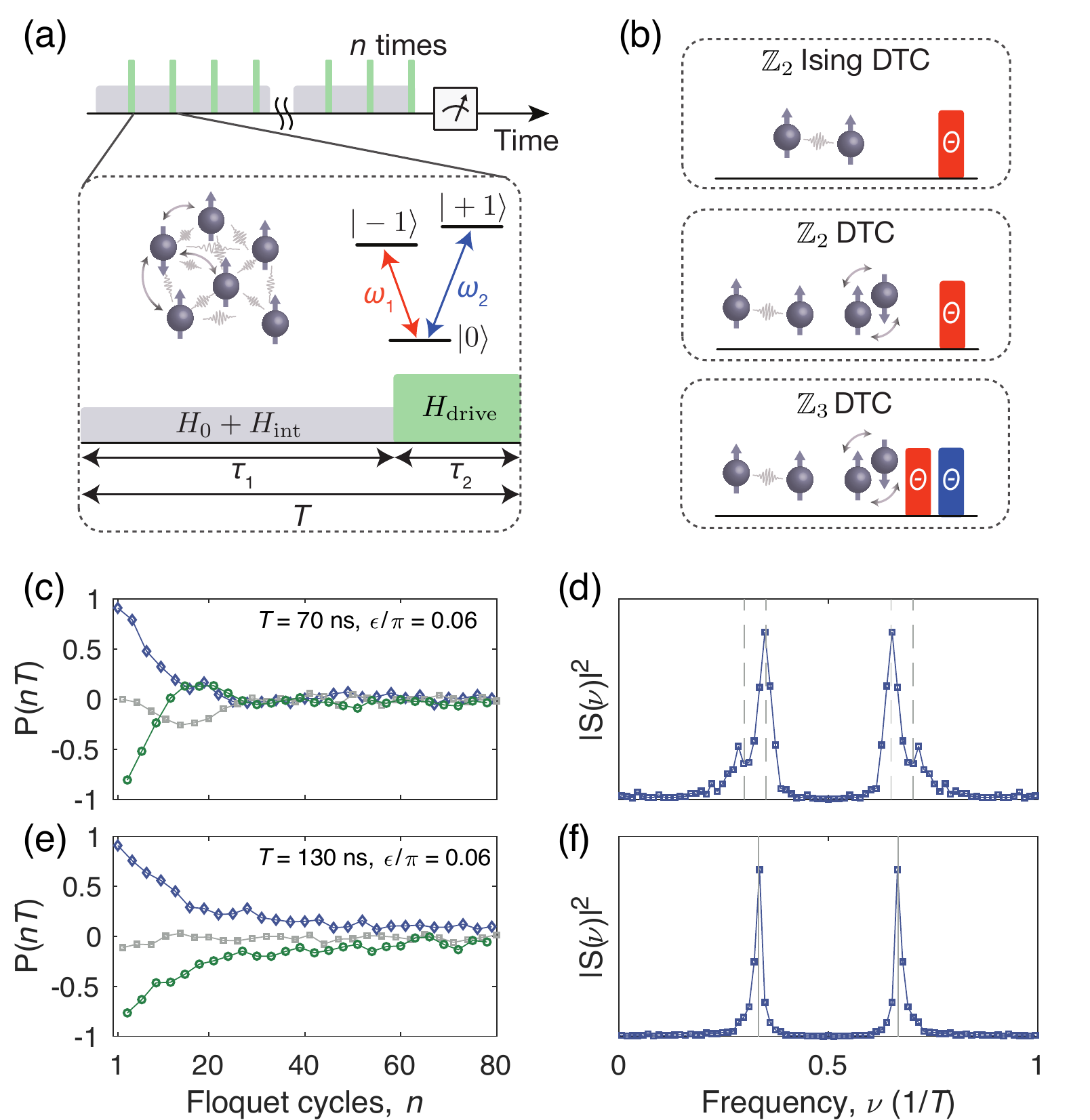}
\caption{\textbf{Experimental system and observation of DTC order.} 
(a) Periodically driven, interacting NV centers. During each Floquet period $T$ (dotted box),  NV centers interact for time $\tau_1$, then experience pulsed microwave rotations of duration $\tau_2$, at resonant frequencies $\omega_1$ or $\omega_2$. After $n$ Floquet cycles, the population difference between $\ket{0}$ and $\ket{-1}$ is measured. (b) Distinct Floquet time evolutions realized: $\Z_2$-Ising, with only Ising interactions, and $\Z_2$ \& $\Z_3$, with both Ising and spin-exchange interactions. (c-f) Representative time traces of the normalized spin polarization $P(nT)$ and Fourier spectra $|S(\nu)|^2$ of the $\Z_3$ DTC order at a perturbation $\epsilon/\pi = 0.06$ with $T$ = 70 ns (c,d)  and 130 ns (e,f). In (c,e), blue, gray, and green points correspond to $P(t)$ at $t\equiv T,2T,3T \pmod{3T}$, respectively.}
\label{fig:fig1}
\end{figure}

{\it Experimental setup.---} Our experiments employ a dense ensemble of nitrogen-vacancy (NV) centers in diamond \cite{Doherty20131}. Each NV center comprises an $S$\,$=$\,$1$ electronic spin with internal states $|m_s$\,$=$\,$0,$\,$\pm1\rangle$, which can be initialized, manipulated, and read out by optical and microwave pulses. The high NV center concentration ($\sim$45\,ppm) in our sample provides strong magnetic dipolar interactions between spins, with interaction strengths significantly exceeding extrinsic decoherence rates \cite{kucsko2016critical,SOM}. Our sample has also various sources of disorder, with an energy scale generally larger than the interaction strength between NV centers. A more detailed characterization can be found in Ref.~\cite{kucsko2016critical, choi2017depolarization,SOM}.

To probe thermalization dynamics, we use pulsed periodic microwave driving to engineer three distinct types of Floquet evolutions, which we denote as $\Z_2$-Ising, $\Z_2$, $\Z_3$. In all cases, a Floquet cycle consists of time evolution under an interacting Hamiltonian for a tunable duration $\tau_1$, followed by strong microwave pulses effecting a global spin rotation:
\begin{align}
U_F^{(a)} = P_\theta^{(a)} \exp[-i H^{(a)} \tau_1],
\end{align}
where $a\in \{\Z_2\textrm{-Ising}, \Z_2, \Z_3\}$, $P_\theta^{(a)}$ is the spin rotation parametrized by a tunable angle $\theta$, and $H^{(a)}$ is an effective Hamiltonian for relevant degrees of freedom of the spin ensemble, containing interaction and disorder terms~[Fig.~\ref{fig:fig1}(a)]. The time durations of $P_\theta^{(a)}$ are fixed at $\tau_2 = 10$ ns ($\Z_2\textrm{-Ising}$ and $\Z_2$) or $\tau_2 = 20$ ns ($\Z_3$), such that the Floquet time period $T = \tau_1 + \tau_2$ is dominated by $\tau_1$. For $\Z_2\textrm{-Ising}$ and $\Z_2$, the microwave excitation $P_\theta^{(a)}$ is resonant with the $\ket{0}\leftrightarrow\ket{-1}$ transition, and these two states form an effective two-level system. For $\Z_3$, $P_\theta^{\Z_3}$ consists of two consecutive pulses, resonant with $\ket{0}\leftrightarrow\ket{\pm1}$ transitions, thereby exploiting all three spin states~[Fig.~\ref{fig:fig1}(b)]. In the ideal case $\theta = \pi$, $P_\pi^{(a)}$ permutes the populations between two (three) spin states such that they return to the original configuration after two (three) cycles. In the following experiments, we introduce systematic perturbations $\epsilon = \theta - \pi$, whose accuracy is limited to about $1\%$ due to spatial inhomogeneity of the applied field and disorder in the system~\cite{SOM}.

The effective spin-spin interactions are different in the three cases. For $\Z_2$ and $\Z_3$, spins interact via natural dipole-dipole interactions, which involve both Ising-type interactions and spin exchange between resonant transitions, e.g. $\ket{0}_i$\,$\otimes$\,$\ket{\pm1}_j$\,$\leftrightarrow$\,$\ket{\pm 1}_i$\,$\otimes$\,$\ket{0}_j$ for spins $i,j$~\cite{SOM}.
For $\Z_2\textrm{-Ising}$, strong transverse microwave driving during $\tau_1$ causes the effective spin-spin interactions in the dressed state basis $\ket{\pm X}$\,$=$\,$(\ket{0}$\,$\pm$\,$\ket{-1})/\sqrt{2}$  to become purely Ising-like~\cite{SOM}. In our experiments, spins are initially polarized along the corresponding quantization axes ($\ket{0}$ for $\Z_2, \Z_3$ and $\ket{+X}$ for $\Z_2$-Ising). After time evolution by the Floquet unitary $U_F^{(a)}$ for $n$ cycles, the remaining polarization $P(nT)$ along the initialization axis is measured via spin-state-dependent fluorescence.

{\it Experimental observations and analyses.---} In all three cases, we observe robust subharmonic responses over a wide range of perturbation strengths $\epsilon$ and Floquet periods $T$. As an example, Fig.~\ref{fig:fig1}(c-f) shows typical time traces of $P(nT)$ and their Fourier spectra $|S(\nu)|^2$ for ${\Z_3}$, for two different $T$ at finite $\epsilon$. For very short $T$, $P(nT)$ shows a modulated decaying signal, and $|S(\nu)|^2$ displays broad sidepeaks at $\epsilon$-dependent locations away from $\nu = 1/3$~[Fig.~\ref{fig:fig1}(c,d)]. For larger $T$, $P(nT)$ instead exhibits long-lived oscillations with a period of three cycles, reflected in a sharp spectral peak pinned at $\nu = 1/3$, indicating that the subharmonic response is stabilized by interactions [Fig.~\ref{fig:fig1}(e,f)]. Generally, we associate the signature of $\Z_m$ DTC order with the presence of $\nu = 1/m$ peaks in the Fourier spectrum that are sharp and robust against perturbations $\epsilon$.
 
\begin{figure}[t]
\includegraphics[width=0.48\textwidth,left]{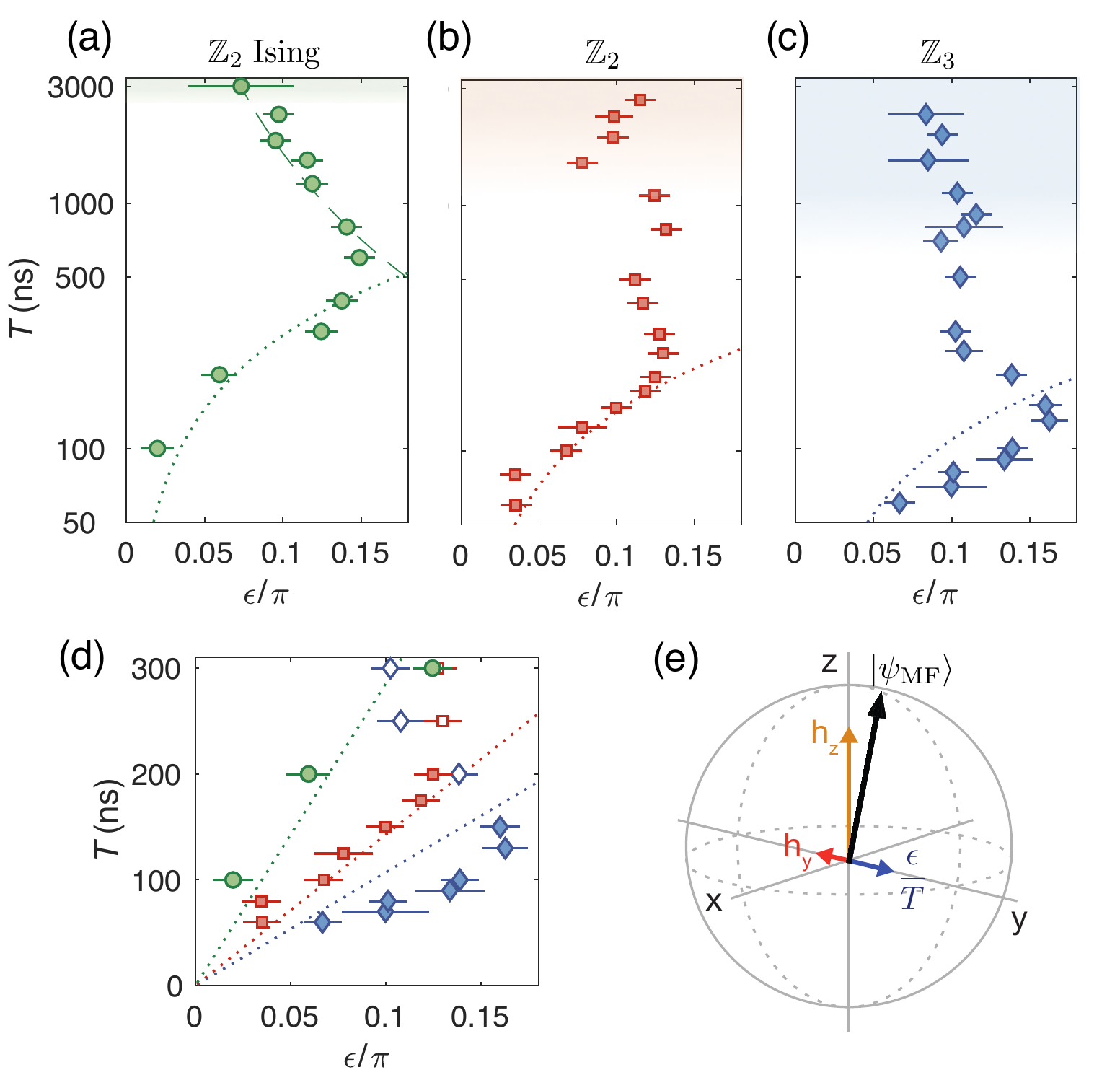}
\caption{\textbf{Phase diagram of DTC order for short drive periods.}
(a-c) Phase diagrams in semi-log scale. The phase boundary (markers) is identified as a crystalline fraction of 10\%. The dotted line indicates the linear phase boundary predicted by a self-consistent mean-field analysis~\cite{SOM}. Shaded areas denote a universal dephasing regime corresponding to Markovian thermalization. In (a), the dashed line represents the theoretical prediction from Ref.~\cite{ho2017critical}. Errorbars denote $95\%$ confidence intervals of the phase boundary~\cite{SOM}. (d) Short-$T$ phase diagram in linear scale (markers as in (a-c)). Open markers indicate points beyond the mean-field regime. (e) Bloch sphere illustrating the screening effect of spin-exchange interactions. $h_z$ and $h_y$ are mean fields arising from Ising and spin-exchange interactions respectively, and $\epsilon/T$ is the perturbing field. The black arrow corresponds to the mean field solution $|\psi_\text{MF}\rangle$.}
\label{fig:fig2}
\end{figure}

To quantitatively probe the stability of DTC order as a function of parameters $\epsilon$ and $T$, we examine the crystalline fraction $f$, defined as the normalized spectral weight at the expected frequency $\nu$\,$=$\,$1/m$ ($m$\,$=$\,$2,3$) in the late time ($n$\,$\geq$\,$40$) dynamics of $P(nT)$, after initial transients in the dynamics have decayed away. For each $T$, we identify the value of $\epsilon$ at $f$\,$=$\,$0.1$ as the phenomenological phase boundary where DTC order is lost [Fig.~\ref{fig:fig2}(a-c)]. Focusing first on short $T$, we find in all three cases that the phase boundaries are linear in the $\epsilon$-$T$ plane, similar to prior observations~\cite{choi2017observation,zhang2017observation,rovny2018observation}. However, closer inspection [Fig.~\ref{fig:fig2}(d)] reveals that DTC order extends to a wider range of $\epsilon$ in $\Z_2$\,\&\,$\Z_3$ than in $\Z_2$-Ising. This is surprising since spin-exchange interactions should intuitively aid thermalization and make DTC order less stable. 

\begin{figure}[t]
\includegraphics[width=.48\textwidth,left]{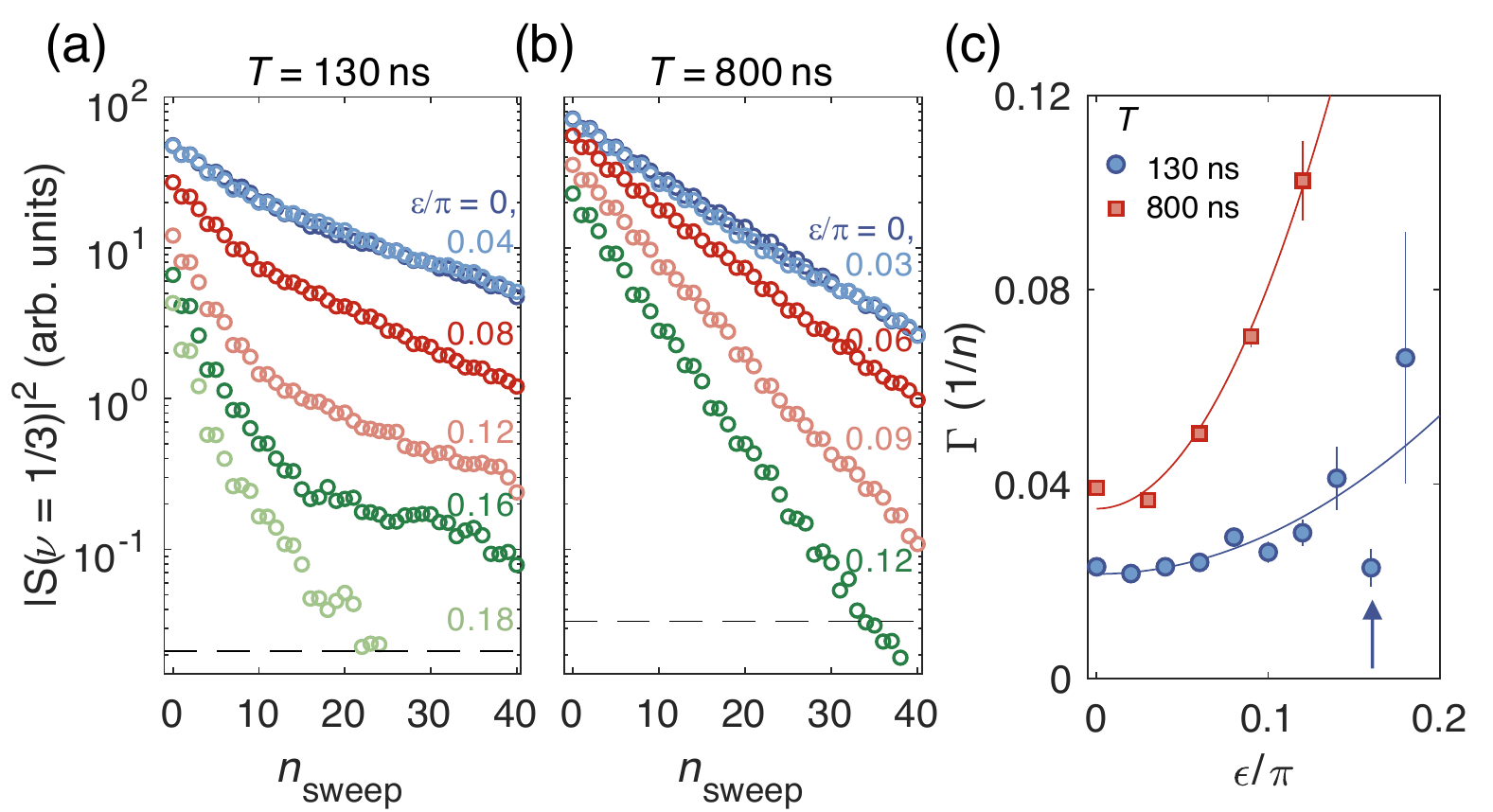}
\caption{\textbf{Long interaction time behavior of DTC order.} (a,b) Temporal decay of $\Z_3$ DTC order $|S(\nu = 1/3)|^2$, for different $\epsilon$, as a function of sweeping window position $n_{\textrm{sweep}}$. Dashed lines denote the noise floor. (c) Late-time decay rate $\Gamma$ as a function of $\epsilon$, with phenomenological quadratic fit. Each data point results from an average over simple exponential fits of $|S(\nu = 1/3)|^2$ starting from $n_{\textrm{sweep}}=15-20$. Error bars denote the statistical error of the fit results. The arrow indicates the mean-field phase boundary.}
\label{fig:fig3}
\end{figure} 
 
From Fig.~\ref{fig:fig2}(a-c), we observe that the linear phase boundaries do not extend indefinitely with increasing $T$, but instead bend inwards, albeit with different shapes between the different Floquet Hamiltonians. To investigate thermalization dynamics in this longer $T$ regime, we examine the decay of DTC order. Specifically, we perform a Fourier transform of $P(nT)$ over a window of cycles $n$\,$\in$\,$[n_\text{sweep},$\,$n_\text{sweep}$\,$+$\,$L$\,$-1]$, where $L$\,$=$\,$36$ is fixed, and extract the subharmonic peak height $\mathcal{S}$\,$=$\,$|S(\nu=1/m)|^2$,\,$(m$\,$=$\,$2,3)$. By sweeping the starting position $n_\text{sweep}$, we produce a time trace of the peak height, which allows us to track how the DTC order decays in time. Fig.~\ref{fig:fig3}(a-b) shows typical decay profiles of $\Z_3$ DTC order, for two different $T$. For $T$ slightly beyond the linear phase boundary regime, the decay exhibits a stretched exponential profile, with late-time decay rates $\Gamma$ nearly independent of $\epsilon$ [Fig.~\ref{fig:fig3}(a)]. In contrast, for long $T$, the decay profile of $\mathcal{S}$ approaches a simple exponential, characteristic of Markovian dynamics [Fig.~\ref{fig:fig3}(b)]. $\Gamma$ also becomes sensitive to $\epsilon$, indicating an instability of the subharmonic signal [Fig.~\ref{fig:fig3}(c)]. We have verified that the other Floquet Hamiltonians also exhibit qualitatively similar changes in behavior of $\mathcal{S}$~\cite{SOM}.

\begin{figure}[t]
\includegraphics[width=.48\textwidth,left]{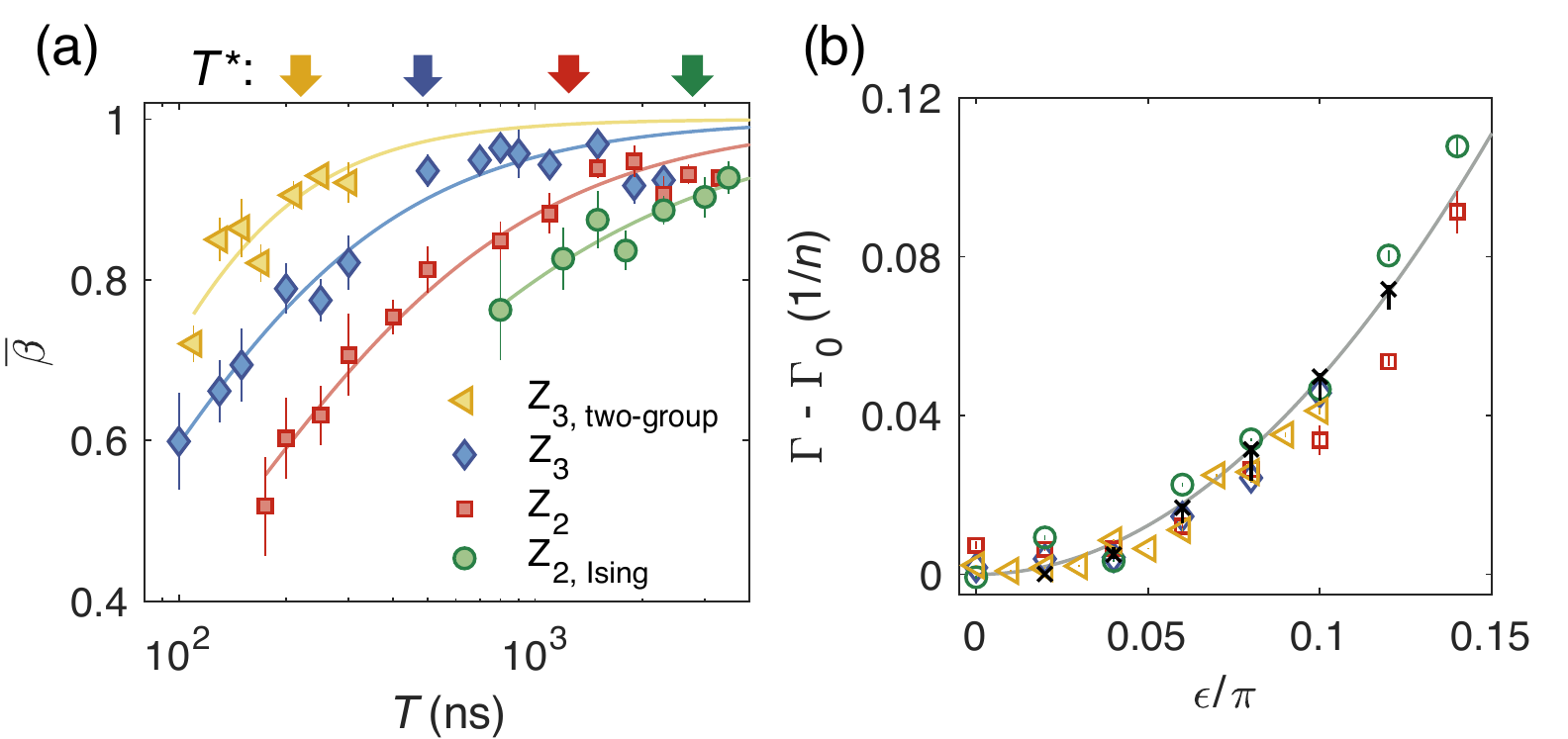}
\caption{\textbf{Universal thermalization dynamics for different Floquet Hamiltonians.} (a) Exponents of the stretched exponential fits versus $T$. Data points denote the average over $\beta$ values extracted at different $\epsilon$, errorbars are the standard deviation of the mean. Lines denote fits to extract the saturation timescale $T^*$ (arrows), identified where $\overline \beta$\,$=$\,$ 0.9$. (b) Late-time decay rate $\Gamma$ as a function of $\epsilon$ (after a global offset $\Gamma_0$ has been subtracted, see~\cite{SOM}) at $T$ = 3.5, 3.5, 2.3 and 0.3 $\mu$s for the $\Z_2$-Ising, $\Z_2$, $\Z_3$, and two-group $\Z_3$ cases, respectively (markers as in (a)). Solid line indicates a dephasing model fit, predicting $\Gamma$\,$=$\,$\epsilon^2/2$\,$+$\,$\Gamma_0$. Error bars as in Fig.~\ref{fig:fig3}(c). Cross markers show numerical results of an infinite-range, interacting spin system with both Ising and spin-exchange interactions~\cite{SOM}. Errorbars in the numerics represent the standard deviation of decay rate distributions for different realizations.}
\label{fig:fig4}
\end{figure}

To quantify the crossover between different decay profiles, we phenomenologically fit $\mathcal{S}$ with a stretched exponential $A\exp[-(n_\text{sweep}/n_T)^\beta]$. For a given $T$, we compute the exponent $\overline \beta$ governing the decay of the stretched exponential, averaged over different $\epsilon$. For all Floquet sequences, we find that $\overline \beta$ increases from $\sim$0.6 (stretched exponential) with increasing $T$, before saturating at 1 (single-exponential), albeit with different saturation timescales $T^*$ [Fig.~\ref{fig:fig4}(a)]. We note there is a slight falling off for very long $T$, which we attribute to convolution effects with the longitudinal spin relaxation ($T_1$) \cite{choi2017depolarization,SOM}. We employ a saturation fit $\overline \beta$\,$=$\,$1/(1+(c_1/T)^{c_2})$ and extract the Floquet period $T^*$ at which $\overline \beta$\,$=$\,$0.9$. Interestingly, $T^*$ coincides with the timescale beyond which $\Gamma$  as a function of $\epsilon$ collapses onto a universal quadratic shape, with curvature approximately equal to 1/2 up to an offset $\Gamma_0$  [Fig.~\ref{fig:fig4}(b)]. Physically, $\Gamma_0$ is attributable to a combination of $T_1$ depolarization of spins and dephasing during the finite rotation pulses \cite{SOM}. For the $\Z_2$-Ising, $\Z_2$, $\Z_3$ cases, we find that $T^*$\,$=$\,$2.64(5), 1.20(5),$\,and\,$0.45(2)$\,$\mu s$ respectively. As expected, $T^*$ is longest for the $\Z_2$-Ising case, indicating that thermalization proceeds slower when only Ising interactions are present.

{\it Discussion.---}
The above observations suggest the existence of three thermalization regimes: a short-$T$ regime where spin-exchange interactions seem to stabilize DTC order, an intermediate-$T$ regime where DTC order persists but is less stable, and a long-$T$, apparently universal regime where subharmonic responses are unstable, decaying at a rate $\Gamma$\,$=$\,$\epsilon^2/2$, and thus cannot be associated with stable DTC order.

To explain our observations in the short-$T$ regime, we turn to a mean-field analysis. When $T$ is sufficiently short compared to the inverse of disorder strength, we can describe the dynamics of the amplitude of $P(nT)$ by an effective, static Hamiltonian $D$ by going into an appropriately chosen moving frame (the so-called toggling frame~\cite{SOM}). Keeping only the lowest order terms in $T$ and $\epsilon$, we obtain
\begin{align}
&D^{\Z_2\textrm{-Ising}} = 
\sum_{ij}\frac{J_{ij}}{r_{ij}^3} S_i^x S_j^x + \frac{\epsilon}{T}\sum_i S_i^y,
\nonumber \\
&D^{\Z_2} = 
\sum_{ij}\frac{J_{ij}}{r_{ij}^3} \left(S_i^x S_j^x + S_i^y S_j^y - S_i^z S_j^z\right) + \frac{\epsilon}{T}\sum_i S_i^y,
\nonumber\\
&D^{\Z_3} =
\sum_{ij} \frac{J_{ij}}{r_{ij}^3} \sum_{ab} \left(\delta_{ab} - \frac{1}{3}\delta_{a\pm1,b}\right) \sigma_i^{ab} \sigma_j^{ba}+ \frac{\epsilon}{3T}  \sum_i R_i,
\nonumber
\end{align}
where $J_{ij},$\,$r_{ij}$ are the orientation-dependent coefficient of dipolar interactions and distance between spins $i,j$, $S_i^\mu$ are spin-1/2 operators for the two levels $\ket{0}$, $\ket{-1}$, $R_i$\,$=$\,$(\sigma_i^{+1,0}$\,$+$\,$\sigma_i^{-1,0}$\,$+$\,$i \sigma_{i}^{+1,-1}$\,$+$\,$h.c.)$, and $\sigma_i^{ab}$\,$=$\,$\ketbra{a}{b}$ with $a,b$\,$\in$\,$\{0,$\,$\pm1\}$. Now, for each Hamiltonian, we seek a self-consistent steady state solution at the mean-field level in the toggling frame, corresponding to a stable subharmonic response in the lab frame. We find that such solutions exist when $\epsilon/T$\,$\leq$\,$a J_{MF}$, where $J_{\textrm{MF}}$ is the total mean-field interaction strength and $a$ equals 1/2, 1, and 4/3 for $\Z_2$-Ising, $\Z_2$, and $\Z_3$ DTC respectively, which yields a linear phase boundary prediction in reasonable agreement with the experimental data [Fig.~\ref{fig:fig2}(d)]. The wider phase boundary in $\Z_2$, $\Z_3$ than $\Z_2$-Ising, can also be understood as a screening effect due to spin-exchange terms having opposite signs to the Ising terms, which partially cancels the perturbing external field under the mean-field approximation (see Fig.~\ref{fig:fig2}(e), ~\cite{SOM}).

The preceding mean-field analysis is expected to break down when $T$ is larger than the inverse of disorder energy scales. Then, resonances due to absorption/emission of energy quanta from/into the drive can occur more readily, giving rise to more thermalization channels. However, the apparent stability of the DTC order even in this regime can be explained---at least in the $\mathbb{Z}_2$-Ising case---by a critical DTC regime~\cite{ho2017critical}, in which the interplay of long-range interactions, dimensionality, disorder and driving leads to critically slow dynamics, and yields a phase boundary narrowing prediction of $\epsilon$\,$\sim$\,$1/\sqrt{T}$. Indeed, Fig.~\ref{fig:fig2}(a) shows that this scaling fits the experimental data extremely well. In contrast, for the $\Z_2$ and $\Z_3$ cases, signatures of stable DTC order remain but are more fragile, existing only in a relatively narrow region. Furthermore, the scaling of the experimentally obtained phase boundaries differs from the $\Z_2$-Ising case.

The observed universal quadratic scaling of decay rates at sufficiently long $T$ can be qualitatively explained by dephasing of individual spins, due to a proliferation of resonances, independent of the details of the thermalizing Hamiltonian. We consider a model \cite{SOM} where each spin is projected onto its quantization axis within one Floquet cycle, leading effectively to Markovian population dynamics, wherein the net ensemble polarization reduces by a factor of $\cos(\epsilon)$ per cycle. This yields a decay rate $\Gamma$\,$=$\,$-\log[\cos(\epsilon)]$\,$\approx$\,$\epsilon^2/2$, which agrees well with the experimental observations [Fig.~\ref{fig:fig4}(b)] upon allowing for an offset $\Gamma_0$ due to external depolarization processes. To probe the origin of dephasing, we perform an additional $\Z_3$ experiment where the effective spin density is doubled~\cite{SOM}, and find that $T^*$ is halved to $0.21(4)$\,$\mu$s [Fig.~\ref{fig:fig4}(a)]. Moreover, the independently estimated dephasing time due to the external bath is much longer than the $T^*$ values \cite{SOM}, strongly suggesting that dephasing dominantly arises from intrinsic, coherent spin-spin interactions. Indeed, exact diagonalization studies of a strongly interacting toy model of $N$ spin-1/2 particles, coupled via all-to-all random interactions $\sum_{ij}$\,$J_{ij}/\sqrt{N}$\,$[\alpha(S_i^x S_j^x$\,$+$\,$S_i^y S_j^y)$\,$-$\,$S_i^z S_j^z]$ with $\alpha$\,$=$\,$0,1$, and which are periodically rotated by an angle $\pi$\,$+$\,$\epsilon$, yield a decay rate $\epsilon^2/2$ of the subharmonic signal for sufficiently long Floquet periods  $J_{ij} T$\,$\gg$\,$1$ [Fig.~\ref{fig:fig4}(b)]. However, we note that the Ising case ($\alpha$\,$=$\,$0$) shows a much slower approach to the Markovian regime in finite-size scaling than the spin-exchange case ($\alpha$\,$=$\,$1$)~\cite{SOM}.

Our observations of the relative stability and distinct scaling of the critical DTC regime in the $\Z_2$-Ising case as well as its long $T^*$ value, indicate important differences in the thermalization dynamics of systems with different types of long-range interactions. This is in broad agreement with recent analytical and numerical studies of thermalization~\cite{safavi2018quantum, burin2006energy, safavi2018quantum, tikhonov2018many}; however, a detailed and better understanding of these differences is a challenging task which deserves further investigation.

{\it Conclusion.---} We have demonstrated that the stability of DTC order can be used to sensitively and quantitatively probe thermalization dynamics of a many-body system. In particular, we have explored how the interplay of disorder, periodic driving, and interactions gives rise to different thermalization regimes. Our results shed light on the mechanisms through which many-body quantum systems approach thermal equilibrium, an important aspect in the quest for full control over quantum matter.

\begin{acknowledgements}
We thank N.~Y.~Yao, K.~X.~Wei, G.~Kucsko for insightful discussions and experimental assistance.
This work was supported in part by CUA, NSSEFF, ARO MURI, Moore Foundation GBMF-4306, Kwanjeong Educational Foundation, Samsung Fellowship, NSF PHY-1506284, NSF DMR-1308435, Japan Society for the Promotion of Science KAKENHI (No.~26246001), EU (FP7, Horizons 2020, ERC), DFG, SNSF, and BMBF.
\end{acknowledgements}

\bibliographystyle{apsrev4-1}
\bibliography{3TDTC_bibtex}

\end{document}


\title{Supplementary Information: Probing quantum thermalization of a disordered dipolar spin ensemble with discrete time-crystalline order}
\affiliation{Department of Physics, Harvard University, Cambridge, Massachusetts 02138, USA}
\affiliation{School of Engineering and Applied Sciences, Harvard University, Cambridge, Massachusetts 02138, USA}
\affiliation{Research Centre for Knowledge Communities, University of Tsukuba, Tsukuba, Ibaraki 305-8550, Japan}
\affiliation{Institut f{\"u}r Quantenoptik, Universit{\"a}t Ulm, 89081 Ulm, Germany}
\affiliation{Takasaki Advanced Radiation Research Institute, National Institutes for Quantum and Radiological Science and Technology, 1233 Watanuki, Takasaki, Gunma 370-1292, Japan}
\affiliation{Sumitomo Electric Industries Ltd., Itami, Hyougo, 664-0016, Japan}
\affiliation{Department of Theoretical Physics, University of Geneva, 1211 Geneva, Switzerland}

\author{Joonhee Choi$^{1,2}$}
\thanks{These authors contributed equally to this work} 
\author{Hengyun Zhou$^{1}$}
\thanks{These authors contributed equally to this work} 
\author{Soonwon Choi$^{1}$} 
\author{Renate Landig$^{1}$}
\author{Wen Wei Ho$^{1}$}
\author{Junichi Isoya$^{3}$}
\author{Fedor Jelezko$^{4}$}
\author{Shinobu Onoda$^{5}$}
\author{Hitoshi Sumiya$^{6}$}
\author{Dmitry A. Abanin$^{7}$}
\author{Mikhail D. Lukin$^{1}$}
\email{lukin@physics.harvard.edu}
\date{\today}


\maketitle

\tableofcontents

\section{Experimental System}
Details of our sample and experimental setup have been described previously in Refs. \cite{kucsko2016critical,choi2017depolarization,choi2017observation}. The nitrogen-vacancy (NV) center in diamond has a spin triplet ground state, labelled as $\ket{m_s = 0, \pm1}$. We use a diamond sample containing a high concentration of NV centers (about 45 ppm), which results in strong magnetic dipolar interactions with a typical interaction strength of $2\pi \times 420\,$kHz. Random positional disorder as well as lattice strain and paramagnetic impurities (P1 centers and $^{13}$C nuclear spins) give rise to a Gaussian-distributed on-site disorder at the NVs with standard deviation $2\pi\times 4.0\,$MHz. 

The diamond sample contains four subgroups of NV centers, each oriented along one of the four different crystallographic axes of the crystal. For the single group measurements, an external magnetic field is applied along one of the crystallographic axes, allowing us to spectrally isolate and independently address the $\ket{0} \leftrightarrow \ket{+1}$ and $\ket{0} \leftrightarrow \ket{-1}$ transitions of the NV group. By using a resonant microwave with different phases, we can apply $\hat{x}$ and $\hat{y}$-rotations to each of the transitions of the spins. At the beginning of each experimental sequence, we initialize the spins into the $|0\rangle$ spin state via 532 nm laser illumination. For high NV density samples, higher laser powers ($> 100~\mu$W) induce a charge instability of negatively charged NVs, leading to a decrease in spin-polarization contrast \cite{choi2017depolarization, giri2017coupled}. To avoid such charge dynamics, we operate at a power of $50\,\mu$W and use a long repolarization duration of $100\,\mu$s. We then apply the desired Floquet pulse sequence. At the end of each experimental sequence, we measure the population difference between the $\ket{-1}$ and $\ket{0}$ states. We also insert a wait time ($\sim$100~$\mu s$) between consecutive sequences to allow the charge states to equilibrate and reduce microwave heating effects. 

For the two-group measurements, we bring two groups of NV centers into resonance by aligning the magnetic field along the bisecting line of the NV axes, in the (1,1,0)-direction relative to the host diamond lattice (the two NV groups are chosen to be in the (1,1,1) and (1,1,-1) directions). These two NV groups will thus have identical transition frequencies (to within $2\pi \times 2$ MHz) and can interact via Ising and spin-exchange interactions, while remaining spectrally isolated from the other NV groups. In addition, the two NV groups are chosen to have similar projections of the microwave driving field, experiencing the same spin rotation for a fixed duration of the microwave pulse. The difference in the rotation angle between the two groups is estimated to be less than 2\%.

\section{Interaction-limited Coherence}
To test whether the decoherence of the dense ensemble is dominated by dipolar interactions among NV centers, we employ an XY8-$N$ dynamical decoupling sequence, with $N$ being the number of repetitions of the XY8 block \cite{staudacher2013nuclear}. Dynamical decoupling sequences such as XY8-$N$ and CPMG have been utilized for various spin systems to investigate the effect of an external spin bath \cite{bar2013solid}. Typically, as $N$ increases, the coherence time of spins is significantly extended due to more effective decoupling from the spin bath, approaching the longitudinal depolarization time $T_1$ \cite{bar2013solid}. However, interactions between like spins are not decoupled by these sequences, and hence if the coherence time is limited by their mutual interactions, then we shall not see a significant extension of coherence time.

\begin{figure}[h]
\includegraphics[width=.7\textwidth]{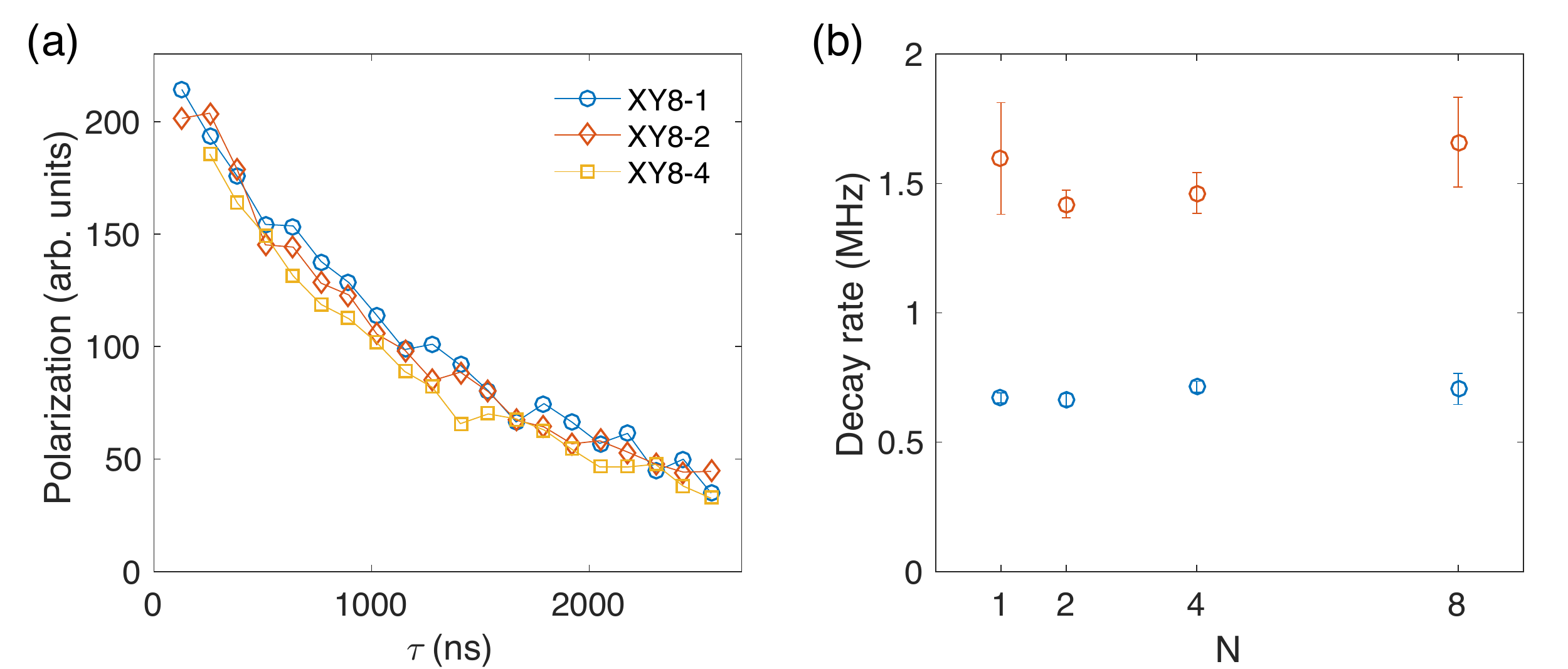}
\caption{\textbf{NV ensemble coherence measured using the XY8-$N$ dynamical decoupling sequence.} (a) Time traces and (b) decay rates of the NV ensemble coherence probed using the XY8-$N$ sequence. In (b), decay rates are extracted from a simple exponential fit to (a). Blue and red data points in (b) correspond to the single- and two-group measurements, respectively. } 
\label{fig:S1}
\end{figure}

As shown in Fig.~\ref{fig:S1}, we measure a decoherence rate $1/T_2$ of 0.7~MHz, independent of the number of repetitions $N$. To reduce finite pulse-width effects in the rotation pulses, we used a short $\pi$ pulse (6 ns). Furthermore, when doubling the spin-spin interaction strength by using two resonant NV groups, we observe a two-fold-enhanced decoherence rate of 1.5~MHz. The observed linear dependence of the decoherence rate on the NV density has also been identified using a spin-echo sequence \cite{kucsko2016critical}. In addition, the correlation time of the external spin bath in our sample was estimated to be $\tau_d \sim40~\mu s$~\cite{kucsko2016critical}, which corresponds to the timescale at which extrinsic noise from the environment can be regarded as Markovian. Indeed, $\tau_d$ is significantly longer than the decoherence times $T_2 \sim1~\mu s$ measured in the dynamical decoupling sequences. This, together with the long correlation time of the bath, strongly suggests that the coherence time is limited by coherent NV-NV interactions.

\section{Implementing the Floquet Hamiltonians}
While for the $\mathbb{Z}_2$-Ising and $\mathbb{Z}_2$ cases, we only address the transition between the $\ket{0}$ and $\ket{-1}$ spin states at resonance frequency $\omega_1$, in the $\mathbb{Z}_3$ case, the spin transition between $\ket{0}$ and $\ket{+1}$ at resonance frequency $\omega_2$ is also addressed (see Fig.~1(b) in the main text). In the $\Z_2$-Ising case, we continuously drive the spins during an interaction period $\tau_1$ (spin-locking) to engineer the spin-spin interaction Hamiltonian. For pulsed rotations, we choose a Rabi frequency of $2\pi\times 50$ MHz, corresponding to a $\pi$ pulse of 10 ns. Perturbations in the global spin rotation are implemented by varying the Rabi frequencies while keeping the pulse duration fixed to $\tau_2$ =10 ns for each pulsed rotation. In the following, we provide detailed explanations for each Floquet Hamiltonian.

\subsection{$\Z_2$-Ising DTC order}
The $\Z_2$-Ising DTC order, exhibiting period-doubled oscillations when the system only has Ising interactions, has already been demonstrated in both ion traps \cite{zhang2017observation} and NV ensembles \cite{choi2017observation}. Due to the large frequency difference between the $\ket{0} \leftrightarrow \ket{-1}$ and $\ket{0} \leftrightarrow \ket{+1}$ transitions in the presence of an external magnetic field, the NV center can be thought of as an effective spin-1/2 system when only one transition is resonantly driven. In our experiments, we use the following pulse sequence to realize $\Z_2$-Ising DTC order. Prior to the Floquet driving, we apply a $\pi/2$ pulse along the $-\hat{y}$ axis to initialize all spins into $\ket{+X} \equiv (\ket{0}+\ket{-1})/\sqrt{2}$. The Floquet period $T$ consists of an interaction duration lasting $\tau_1$ and a global spin rotation of length $\tau_2$. During $\tau_1$, we suppress spin-exchange interactions by continuously driving the spins with a Rabi frequency of $2\pi \times\, 41.7$~MHz along the $\hat{x}$ axis. Afterwards, we perform a global spin rotation by an angle $\theta=\pi+\epsilon$ around the $\hat{y}$ axis. After $n$ repetitions of the Floquet period, a $\pi/2$ pulse along the $\hat{y}$ axis is applied to read out the spin polarization along the $\hat{x}$ axis. In the rotating frame, the effective Hamiltonian for the $\Z_2$-Ising case can be described as 
\begin{align} \label{eq:HamZ2Ising}
H(t) = \sum_i (\Omega_x (t)  S_i^x + \Omega_y (t)  S_i^y +\Delta_i S_i^z) + \sum_{ij} \frac{J_{ij}}{r_{ij}^3} S_i^x S_j^x
\end{align}
where $\Omega_x(t)$ and $\Omega_y(t)$ are the Rabi frequencies for spin-locking and rotation pulses, and are turned on only during the interaction and rotation parts in each period, respectively. Here, $J_{ij}$ is the orientation-dependent interaction strength and $r_{ij}$ is the distance between the NV centers at site $i$ and $j$, respectively, $\Delta_i$ is the on-site disorder field at site $i$, and $\vec{S} = \{S^x,S^y,S^z\}$ are the spin-1/2 operators. It has been shown that the long-time evolution of a driven system is governed by an average Hamiltonian for times exponentially long in the driving frequency \cite{abanin2017,machado2017exponentially}. Applying average Hamiltonian theory in the toggling frame (Eq.~(2) in the main text) to Eq.~(\ref{eq:HamZ2Ising}) transforms it into
\begin{align} \label{eq:DZ2Ising}
D^{\Z_2,\text{Ising}} \simeq \sum_i \frac{J_{ij}}{r_{ij}^3} S_i^x S_{j}^x  + \frac{\epsilon}{T} \sum_i  S_i^y.
\end{align}
We note that the average Hamiltonian for the $\Z_2$-Ising case is equivalent to a long-range transverse-field Ising model. Here, $\epsilon=\theta-\pi$ is the perturbation due to imperfect rotations away from $\pi$. Intuitively, the Ising interaction in Eq.~(\ref{eq:DZ2Ising}) gives rise to a long-range spatiotemporal correlation of spins along the $\hat{x}$ axis when the total mean-field interaction strength $J_{\text{MF}} = \langle \sum_i J_{ij}/r_{ij}^3 \rangle$ dominates over the perturbation $\epsilon/T$. $\langle \cdots \rangle$ denotes averaging over different positional configurations of disordered spins.

\subsection{$\Z_2$ DTC order}
We are also interested in probing period-doubled oscillations in the presence of spin-exchange interactions, which we here denote as $\Z_2$ DTC order to distinguish from the preceding case with Ising terms only. If we work along the $\hat{z}$ axis in the bare basis, without any microwave driving during the interaction period $\tau_1$, both spin-exchange and Ising interactions will be present in the effective Hamiltonian. In the experiment, we first initialize the spins into the $\ket{0}$ spin state. During $\tau_1$, spins evolve under the bare dipolar Hamiltonian, which includes both Ising and spin-exchange interactions. After $\tau_1$, the spins are all rotated by an angle $\theta = \pi + \epsilon$ around the $\hat{y}$ axis in the subspace spanned by $\ket{0}$ and $\ket{-1}$. After $n$ repetitions of the Floquet period, we read out the spin polarization along the $\hat{z}$ axis. Treating the NV center as an effective spin-1/2 system, the effective Hamiltonian for the $\Z_2$ DTC order (in the rotating frame) can be expressed as follows:
\begin{align} \label{eq:HamZ2}
H(t) = \sum_i \Omega_y (t) S_i^y +\Delta_i S_i^z + \sum_{ij} \frac{J_{ij}}{r_{ij}^3} \left( S_i^x S_j^x + S_i^y S_j^y - S_i^z S_j^z \right),
\end{align}
where $S_i^x S_j^x + S_i^y S_j^y = \frac{1}{2}(S_i^+ S_j^- + S_i^- S_j^+)$ is the spin-exchange interaction term that leads to flip-flop processes between the spins at sites $i$ and $j$. Here, $S^{\pm} = S^x \pm iS^y$. To capture the long-time behavior of the $\Z_2$ DTC order, we apply average Hamiltonian theory to Eq.~(\ref{eq:HamZ2}), which yields 
\begin{align} \label{eq:DZ2}
D^{\Z_2} \simeq \sum_i \frac{J_{ij}}{r_{ij}^3} \left( S_i^x S_j^x + S_i^y S_j^y - S_i^z S_j^z \right)  + \frac{\epsilon}{T} \sum_i  S_i^y.
\end{align}
Compared to the transverse-field Ising model described above, Eq.~(\ref{eq:DZ2}) additionally contains spin-exchange interactions. Interestingly, the mean-field analysis detailed in the section, {\it{Mean-Field Description for Short Interaction Time Regime}}, reveals that in the fast Floquet driving limit, i.e., $2\pi/T \gg J_{ij} / r_{ij}^3$, the DTC phase becomes more robust against perturbations due to the presence of the spin-exchange interaction. As depicted in Fig.~2(d) of the main text, we attribute this behavior to the creation of an additional mean-field by the spin-exchange interactions, which counteracts and reduces the perturbation strength $\epsilon$. 

\subsection{$\Z_3$ DTC order}
To observe $\Z_3$ DTC order, in which the system exhibits period-tripled oscillations, we work in the bare basis while utilizing all three spin states $\ket{m_s = 0, \pm1}$. We start with all spins polarized into the $\ket{0}$ state and evolve under the bare spin-1 dipolar Hamiltonian for a duration $\tau_1$. Subsequently, we apply two resonant microwave pulses, first on the transition $\ket{0} \leftrightarrow \ket{-1}$ and then on the transition $\ket{0} \leftrightarrow \ket{+1}$. The two consecutive rotation pulses are separated by 1 ns to avoid microwave interference. The combination of these operations defines a Floquet cycle with period $T$. After $n$ repetitions of the Floquet period, we measure the population difference between the $\ket{0}$ and $\ket{-1}$ spin states. When each of the applied pulses corresponds to a perfect $\pi$-pulse, this sequence realizes a cyclic transition with $\Z_3$ symmetry. However, this discrete symmetry is explicitly broken by imperfect rotations whose angle deviates from $\theta = \pi$. Considering the full spin-1 nature of the NV centers, we describe the effective Hamiltonian for the $\Z_3$ DTC order as
\begin{align}
H(t) = &\sum_i \Omega^- (t) (\sigma_i^{-1,0}+\sigma_i^{0,-1}) + \Omega^+ (t) (\sigma_i^{+1,0}+\sigma_i^{0,+1}) + \Delta^-_i \sigma_i^{-1,-1} + \Delta^+_i \sigma_i^{+1,+1} \\
&+\sum_{ij} \frac{J_{ij}}{r_{ij}^3} \label{eq:Z3DTC}
\left[
-\frac{\sigma_i^{+1,0}\sigma_j^{0,+1}+\sigma_i^{-1,0}\sigma_j^{0,-1}+h.c.}{2}+ (\sigma_i^{+1,+1}-\sigma_i^{-1,-1})(\sigma_j^{+1,+1}-\sigma_j^{-1,-1})
\right],
\end{align}
where $\Omega^-(t)$ and $\Omega^+(t)$ are the Rabi frequencies for the rotation pulses acting on the lower $\ket{0} \leftrightarrow \ket{-1}$ and upper $\ket{0} \leftrightarrow \ket{+1}$ transitions, respectively, and are turned on separately only during the rotation period $\tau_2$. Here, $\sigma^{ab}_i$ = $\ket{a}_i\bra{b}$, and $\Delta^-_i$ and $\Delta^+_i$ are the on-site disorders for the lower and upper transitions of the spin at site $i$.  
In our experiment, we choose a common Rabi frequency $|\Omega^-|$ = $|\Omega^+|$ and tune its amplitude to control the spin rotation angle $\theta$. For fast Floquet driving, the evolution is governed by the following average Hamiltonian (in the toggling frame):
\begin{align}
D^{\Z_3} \simeq  
\sum_{ij} \frac{J_{ij}}{r_{ij}^3} 
\sum_{ab} \left(\delta_{ab} - \frac{1}{3}\delta_{a\pm1,b} \right) \sigma_i^{ab}\sigma_j^{ba} 
+ \frac{\epsilon}{3T}  \sum_i \left(\sigma_i^{+1,0} + \sigma_i^{-1,0} + i \sigma_{i}^{+1,-1} + h.c.\right),
\end{align}
where, in the first term, $\sum_{ab} \delta_{ab} (\sigma_i^{ab}\sigma_j^{ba}) $ and $\sum_{ab} \delta_{a\pm1, b} (\sigma_i^{ab}\sigma_j^{ba})$ represent the Ising and spin-exchange interactions, respectively. Similar to the $\Z_2$ DTC order, the presence of flip-flop processes in the $\Z_3$ DTC sequence also leads to a reduction in the effective perturbation strength, making the $\nu = 1/3$ DTC order more robust. Further details are provided in the section, {\it{Mean-Field Description for Short Interaction Time Regime}}.

\subsection{Two-group $\Z_3$ DTC order}
By tuning the orientation of an externally applied magnetic field, we spectrally overlap two NV groups to within $(2\pi)\times2$ MHz. For the implementation of the Floquet Hamiltonian, we follow the same protocol used for the single-group $\Z_3$ DTC order. However, it is noteworthy that the interaction Hamiltonian for the two-group $\Z_3$ DTC order is not perfectly identical to Eq.~(\ref{eq:Z3DTC}) owing to different crystallographic axes of the two NV groups. Depending on the spatial orientation of the two spins belonging to different groups, the relative strength as well as sign between the Ising and spin-exchange interactions can be different; on average however, the relative sign between Ising and spin-exchange terms is still negative. Despite such differences, the two-group $\Z_3$ DTC order also shows universal thermalizing dynamics in the long interaction time regime, consistent with the single-group DTC measurements (see the section below, {\it Universality in Long Interaction Time Regime}).

\section{Phase Boundary Extraction}
\begin{figure}[h]
\includegraphics[width=.7\textwidth]{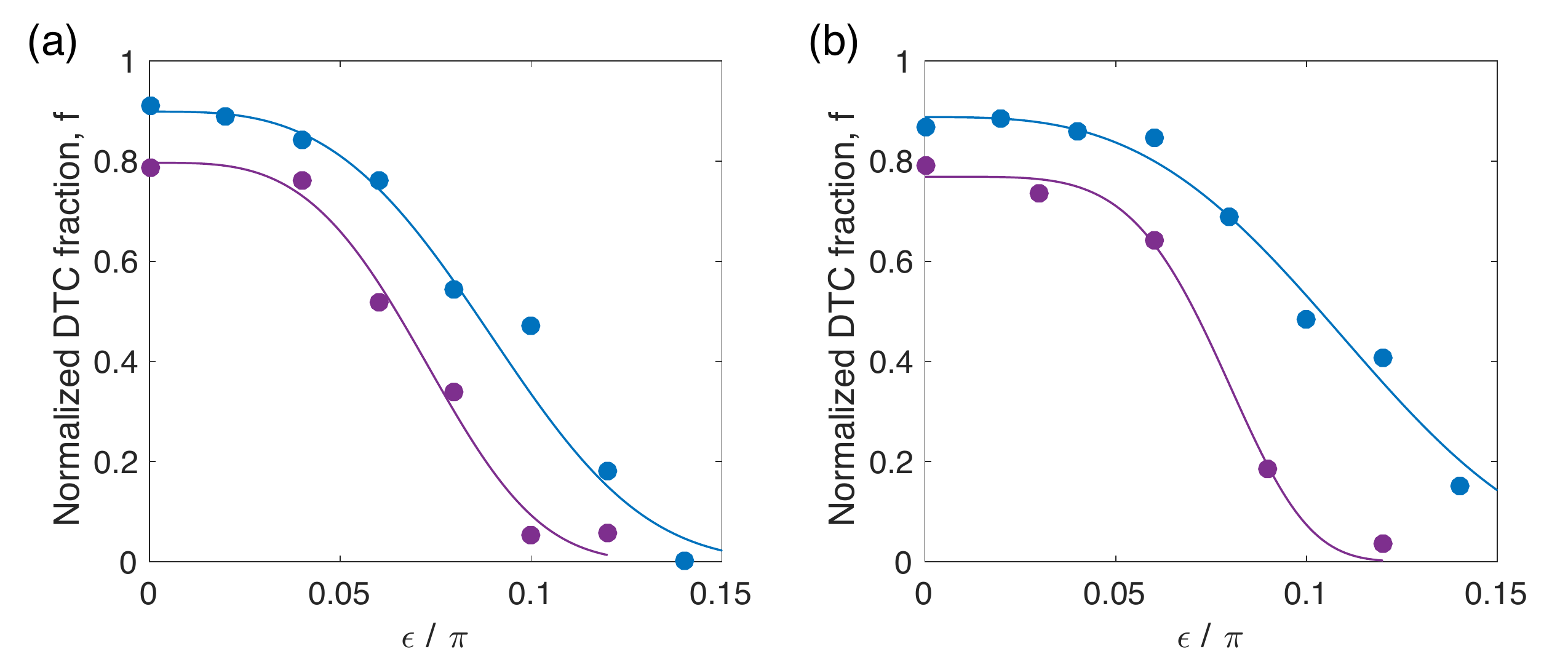}
\caption{\textbf{Late-time crystalline fraction analysis.} (a) $\Z_2$ case with $T$ = 200 ns (blue) and $T$ = 1900 ns (purple). (b) $\Z_3$ case with $T$ = 130 ns (blue) and $T$ = 800 ns (purple). In (a,b), solid lines represent the super-Gaussian fits. } 
\label{fig:S2}
\end{figure}

To experimentally extract the DTC phase boundary, we follow the method developed in our previous manuscript, Ref.~\cite{choi2017observation}. In short, we quantify the DTC order by using the crystalline fraction, defined as $f=|S(\nu=1/2)|^2/\sum_{\nu}|S(\nu)|^2$ and $f=2|S(\nu=1/3)|^2/\sum_{\nu}|S(\nu)|^2$ for the $\Z_2$ and $\Z_3$ DTC order, respectively. For the $\Z_3$ case, the factor of 2 in $f$ takes into account the mirror symmetry of the Fourier-transform spectra with respect to $\nu = 1/2$. At each Floquet period $T$, we examine $f$ as a function of $\epsilon$. Fig.~\ref{fig:S2} shows representative data together with a phenomenological super-Gaussian fit function 
\begin{align}
f_{T} (\epsilon) = f_{T}^{\text{max}}\exp  \left[ -\frac{1}{2} \left( \frac{|\epsilon - \epsilon_0|}{\sigma} \right)^p  \right]  ,
\end{align}
where $\epsilon_0$, $\sigma$, $p$ are the central position, characteristic width, and power of the super-Gaussian fit, and $f_T^\text{max}$ is the maximum value of the crystalline fraction for a given Floquet period $T$. As $p$ increases, the functional profile becomes flat when $|\epsilon - \epsilon_0| < \sigma$ and rolls off sharply when $|\epsilon - \epsilon_0| > \sigma$. We identify the phase boundary at a given $T$ as the value of $\epsilon$ for which $f_T=0.1$ (see Fig. 2(a-c) in the main text). Horizontal errorbars on the phase boundary correspond to a 95\% confidence interval from the fit.

\section{Mean-Field Description for Short Interaction Time Regime}
In this section, we provide a simple description of the dynamics in the short interaction time regime (or equivalently, fast Floquet drive regime), where the Floquet drive energy scale $\omega_0 = (2\pi)/T$ is larger than the on-site disorder strength as well as typical interaction strengths in the system. In such a case, the system cannot effectively absorb or emit energy from/to the drive, and the dynamics can be well-captured by an average Hamiltonian model \cite{haerberlen1968coherent}. 

In the following, we analyze the system under the framework of mean-field theory to understand the conditions for the emergence of an ordered stationary state, which will aid in understanding the phase boundary at short interaction times. Note that our system is favorable to such mean-field analysis as it has long-range interactions among spins in high dimensionality.

First, we consider the case of $\Z_2$-Ising, where there are only Ising interactions between the spins, arriving at results that are consistent with previous analysis \cite{choi2017observation}. The average Hamiltonian in two Floquet cycles, with imperfect rotation angle $\pi+\epsilon$, is given by
\begin{align}
D^{\Z_2,\text{Ising}} &\simeq \sum_{ij}\frac{J_{ij}}{r_{ij}^3}S_i^zS_j^z+\frac{\epsilon}{T}\sum_iS_i^y \\
&=\sum_{ij}\frac{J_{ij}}{r_{ij}^3}\sum_{\mu\nu}C_{\mu\nu}S_i^\mu\otimes S_j^\nu+\sum_{i\mu} h_\mu S_i^\mu,
\end{align}
with $C_{\mu\nu}$ being a diagonal matrix with diagonal elements $(0,0,1)$ and $h_\mu=(0,\frac{\epsilon}{T},0)$ (we have permuted the basis definition for ease of comparison with the $\Z_2$ case).
Here $S_i^\mu$ ($\mu\in\{x,y,z\}$) are the spin-1/2 operators acting on the two-level system of interest, and in the final step we have written the expression in a more general form for ease of analysis under other interparticle interactions. Note that the static on-site disorder has been echoed out to leading order in this effective Hamiltonian owing to periodic rotation pulses in the lab frame.

Under the mean-field approximation, we replace two-body interactions with single-body terms by taking the expectation values of one of the spins: $S_i^\mu\otimes S_j^\nu \mapsto S_i^\mu\langle S_j^\nu\rangle$. We then self-consistently evaluate the expectation value $\langle S_j^\nu\rangle$ by plugging in the corresponding values calculated from spin $i$. Replacing the disordered interaction strength $J_{ij}/r_{ij}^3$ by a total mean-field interaction strength $J_{\text{MF}} = \langle \sum_i J_{ij}/r_{ij}^3 \rangle$, we obtain the mean-field Hamiltonian
\begin{align}
H_{\text{MF}}=\sum_\mu \left(J_{\text{MF}}\sum_\nu C_{\mu\nu}\langle S^\nu\rangle + h_\mu\right)S^\mu,
\end{align}
with $C_{\mu\nu}$ and $h_\mu$ given above. With $H_{\text{MF}}$, we seek a stationary solution for the density matrix of a spin $\rho=\frac{I_{2 \times 2}}{2}+\sum_\mu\rho_\mu S^\mu$ under the dynamics defined by $H_{\text{MF}}$:
\begin{align}\label{eq:MFEq}
\dot{\rho}=i[\rho,H_{\text{MF}}]=0,
\end{align}
subject to the self-consistency condition
\begin{align}\label{eq:SelfConsistence}
\langle S^\mu\rangle=\textrm{tr}\left[S^\mu \rho\right]=\rho_\mu/2.
\end{align}
Solving this set of equations yields two solutions, only one of which allows a nonzero expectation value of $\langle S^z\rangle$. This solution imposes $\rho_x=\frac{2(\epsilon/T)}{J_{\text{MF}}}$ and $\rho_y=0$. Given the restriction
\begin{align}\label{eq:Norm}
\textrm{tr}\left[\rho^2\right]\leq 1
\end{align}
on the density matrix and using the relation $\textrm{tr}[S^\mu S^\nu]=\delta_{\mu\nu}/2$, we find that $\sum_\mu \rho_\mu^2\leq 1$. Thus, a stationary self-consistent mean-field solution exists only when $\epsilon/T \leq J_{\text{MF}}/2$.

The mean-field approach implies a linear phase boundary at short Floquet periods, with a slope given by the interaction strength of the system with prefactor $1/2$; this is consistent with the procedure employed in previous papers, where the same result was derived by examining the rotations of Floquet eigenstates under self-consistent mean-fields. An intuitive understanding of the derivation presented above is that it allows us to find a self-consistent product-state ansatz, for which the rotation induced by interaction with other spins compensates the imperfect rotations imposed by the Floquet drive. Therefore, the robust DTC response can be understood as the existence of a period-doubled trajectory that is stable against perturbations. As we shall see below, a similar intuition applies to the case with spin-exchange interactions as well.

Now, we consider the case of $\Z_2$ with Ising as well as spin-exchange interactions. The average Hamiltonian in two Floquet cycles, with imperfect rotation angle $\pi+\epsilon$, is given by
\begin{align}
D^{\Z_2} &\simeq \sum_{ij}\frac{J_{ij}}{r_{ij}^3}\left(-S_i^xS_j^x - S_i^yS_j^y + S_i^zS_j^z\right)+\frac{\epsilon}{T}\sum_iS_i^y \\
&=\sum_{ij}\frac{J_{ij}}{r_{ij}^3}\sum_{\mu\nu}C_{\mu\nu}S_i^\mu\otimes S_j^\nu+\sum_{i\mu} h_\mu S_i^\mu,
\end{align}
where now the coefficient matrices are: $C_{\mu\nu}$ diagonal with elements $(-1,-1,1)$ and $h_\mu=(0,\frac{\epsilon}{T},0)$. Repeating the same procedure using Eq.~(\ref{eq:MFEq}) and Eq.~(\ref{eq:SelfConsistence}), we find that the solution with nonzero expectation value of $\langle S^z\rangle$ is given by $\rho_x=0$ and $\rho_y=\frac{(\epsilon/T)}{J_{\text{MF}}}$. Eq.~(\ref{eq:Norm}) gives the normalization condition for a stationary self-consistent mean-field solution as $\epsilon /T \leq J_{\text{MF}}$. Thus, we expect the phase boundary at short interaction periods, in the presence of spin-exchange interactions, to remain linear, but with a two-fold increase in slope that results in a phase boundary width twice as wide as the case of Ising interactions.

This result can also be intuitively understood by examining dynamics on the Bloch sphere (see Fig.~2(d) in the main text); in the self-consistent solution above, the spins develop a nonzero expectation value along the positive $\hat{y}$ axis, which in turn generates a mean-field along the negative $\hat{y}$ axis due to the spin-exchange terms in the Hamiltonian. This mean-field along the $\hat{y}$-direction counteracts the applied external perturbation $\epsilon/T$, resulting in a smaller effective perturbation. Therefore, the DTC order becomes more robust and the phase boundary expands to a larger $\epsilon$ value compared to the case where there are only Ising interactions.

We note that the additional stabilizing effect arising from the spin-exchange interactions is present when the initial state is polarized and the spin-exchange terms have opposite signs to the Ising terms, e.g. $J_{ij}(-[S^xS^x+S^yS^y]+S^zS^z)$, regardless of the overall sign of the interaction.  Geometrically, this can be seen by considering how a pair of interacting spins evolves under the action of an applied perturbation (a similar intuition can be generalized to clusters of spins). In the absence of perturbations $\epsilon$ and with a polarized initial state, a stationary solution occurs when the spins are pointing in the same direction along the $\hat{z}$-axis; depending on the global sign of the mutual interaction $sgn(\sum_j J_{ij})$, this will correspond to each spin being either aligned or anti-aligned to its local field. When a perturbation is applied, the spin direction will adiabatically follow the total field. This means that in the case of a positive interaction $\sum_jJ_{ij}>0$ (aligned), the spins will tilt in the same direction as the applied field, while for a negative interaction $\sum_jJ_{ij}<0$ (anti-aligned), the spins will tilt in the opposite direction. The expectation value of the spin vector thus depends on the sign of $\sum_jJ_{ij}$, and hence the mean-field acting on each spin, which has an additional factor of $J_{ij}$, will always have the correct sign to counteract the applied perturbation. Therefore, the spin-exchange terms will lead to a reduction in perturbations only when they have opposite signs to the Ising terms.

Finally, we perform a similar derivation for the case of $\Z_3$. As previously derived \cite{choi2017observation}, the effective Hamiltonian over three Floquet periods is given by
\begin{align}
D^{\Z_3} &\simeq   
\sum_{ij} \frac{J_{ij}}{r_{ij}^3} \sum_{ab} \left(\delta_{aa} - \frac{1}{3}\delta_{a\pm1,b} \right)
\sigma_i^{ab}\sigma_j^{ba} + \frac{\epsilon}{3T}  \sum_i \left(\sigma_i^{+1,0} + \sigma_i^{-1,0} + i \sigma_{i}^{+1,-1} + h.c.\right)
 \\
&=\sum_{ij} \frac{J_{ij}}{r_{ij}^3} 
 \sum_{\mu \nu} C_{\mu \nu} \lambda^\mu_i \otimes \lambda^\nu_j + \sum_{i\mu} h_\mu \lambda_i^\mu,
\end{align}
where $\sigma^{ab}_i = |a\rangle_i\langle b|$ for a spin at site $i$, and in the last line we have re-expressed the Hamiltonian in the orthonormal Gell-Mann matrix basis $\lambda^\mu$, which satisfies $\textrm{tr}[\lambda^\mu\lambda^\nu]=2\delta_{\mu\nu}$. The coefficient matrices are: $C_{\mu\nu}$ is a diagonal matrix, with diagonal elements $C_{\mu\mu}=(-\frac{1}{6},-\frac{1}{6},-\frac{1}{6},-\frac{1}{6},-\frac{1}{6},-\frac{1}{6},\frac{1}{2},\frac{1}{2})$, and $h = (\frac{\epsilon}{3T},\frac{\epsilon}{3T},0,0,0,-\frac{\epsilon}{3T},0,0)$. Our convention is such that the last two Gell-Mann matrices correspond to nonzero population imbalances between the different spin states. 

Working in the spin-1 manifold, we write the density matrix as $\rho=\frac{I_{3 \times 3}}{3}+\sum_\mu \rho_\mu\lambda^\mu$. Repeating the same procedure using the spin-1 equivalents of Eq.~(\ref{eq:MFEq}) and Eq.~(\ref{eq:SelfConsistence}), we find solutions to the self-consistent equations. Of the solutions to this set of equations, the only physically-relevant (normalizable) solution with nonzero expectation value in the population imbalance (in the limit of $\epsilon\rightarrow0$) is given by
\begin{align}
\rho_1=\rho_2=-\rho_6&=\frac{(\epsilon/T)}{4J_{\text{MF}}}, \\
 \rho_3=\rho_4=\rho_5&=0,
\end{align}
and $\rho_7$, $\rho_8$ are arbitrary numbers depending on the initial conditions and satisfying the normalization requirements. Using the trace orthonormality of Gell-Mann matrices, we find that Eq.~(\ref{eq:Norm}) imposes the constraint $\sum_\mu\rho_\mu^2\leq 1/3$, which implies that a self-consistent solution exists when $\epsilon/T \leq 4J_{\text{MF}}/3$. This shows that the phase boundary is expected to be even wider in the case of $\Z_3$ with Ising and spin-exchange interactions, compared to the preceding two cases.

In conclusion, we have derived the conditions for which a stationary self-consistent mean-field solution exists at the average Hamiltonian level, for each of the different DTC realizations and their associated interaction Hamiltonians. To summarize, we have found that
\begin{align}
\frac{(\epsilon/T)}{J_{\text{MF}}}\leq \begin{cases}
\frac{1}{2},\quad &\textrm{ $\Z_2$ DTC with Ising interactions;}\\
1,\quad &\textrm{ $\Z_2$ DTC with Ising and spin-exchange interactions;}\\
\frac{4}{3},\quad &\textrm{ $\Z_3$ DTC with Ising and spin-exchange interactions.}\\
\end{cases}
\end{align}
As shown in Fig.~2 of the main text, these theoretical predictions are in fairly good agreement with the experimental phase boundaries. We note that the total mean-field interaction strength $J_{MF} = 2\pi \times$ 350 kHz is consistent with the independently extracted typical interaction strength of the system $\sim 2\pi \times 105 $ kHz \cite{kucsko2016critical}, as can be seen from the phase diagrams in Ref.~\cite{choi2017observation,ho2017}, which used Monte Carlo simulations to estimate the total mean-field on each individual spin due to the combination of all other spins.

\begin{figure}[h]
\includegraphics[width=0.9\textwidth]{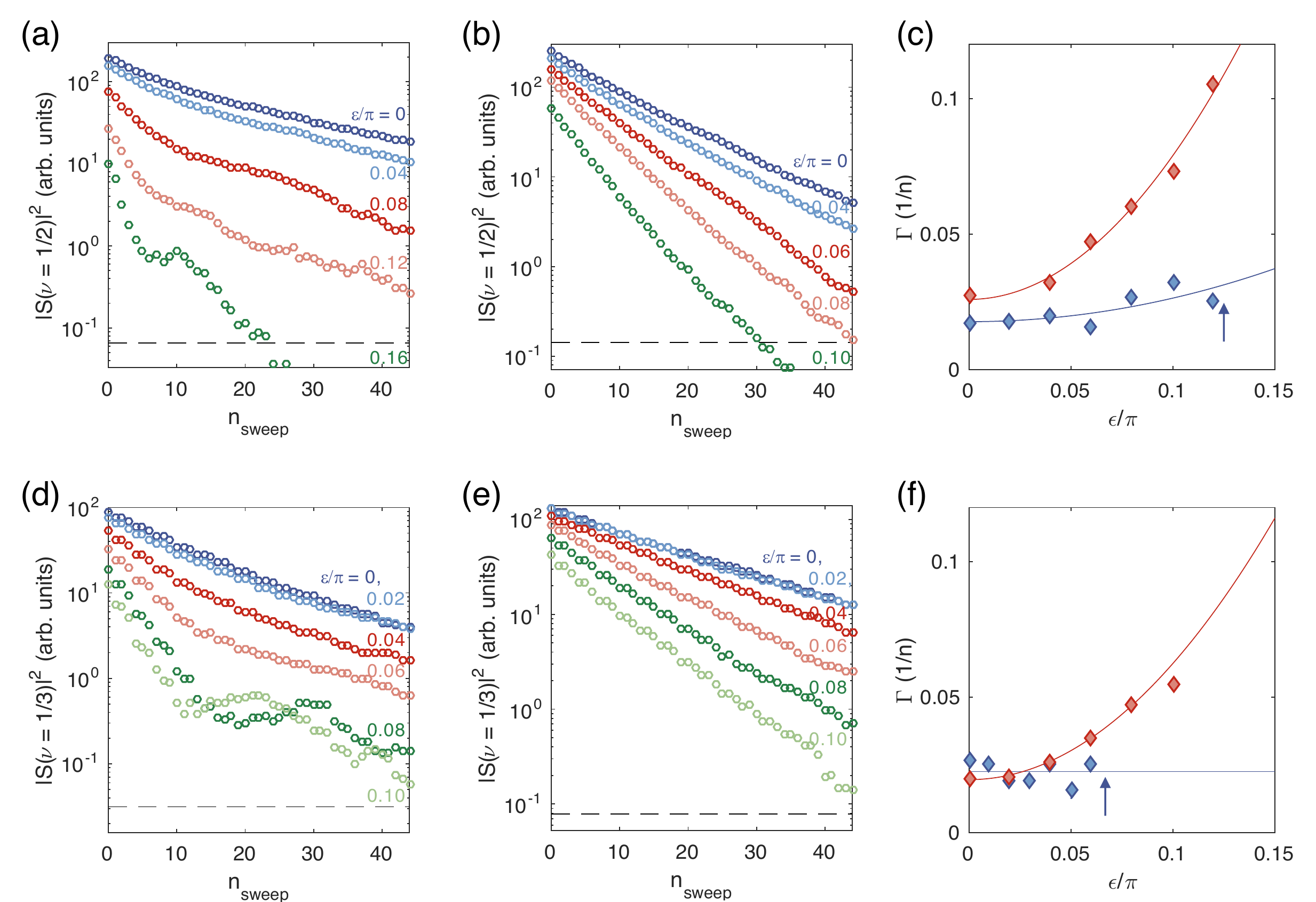}
\caption{\textbf{Probing decay rates of DTC order at short and long interaction times.} Representative time traces and late-time decay rates of (a-c) $\Z_2$ and (d-f) two-group $\Z_3$ DTC peak heights. In (a,b,d,e), we present the time traces at (a) $T$ = 200 ns and (b) $T$ = 1900 ns for the $\Z_2$, and (d) $T$ = 70 ns and (e) $T$ = 250 ns for the two-group $\Z_3$, respectively, and the dashed gray lines denote noise floors. In (c,f), the late-time decay rates of DTC order as a function of perturbation strength are compared between short and long interaction times for each case. In (c), blue and red data correspond to $T$ = 200 ns and $T$ = 1900 ns, respectively, while in (f), they correspond to $T$ = 70 ns and $T$ = 250 ns. Arrows indicate the phase boundary for short interaction times where either the crystalline fraction drops below 10\% or sidepeaks away from the DTC peak position are predominant. Solid lines are phenomenological quadratic fits, $\Gamma(\epsilon) = \Gamma_0 + a\epsilon^2$. } 
\label{fig:S3}
\end{figure}

\section{Lifetime Analysis}
The late-time decay rate $\Gamma$ of the DTC peak height is extracted in two steps. First, the height of the $\nu=1/2$ ($\Z_2$) or $\nu=1/3$ ($\Z_3$) peak is determined by short-time Fourier transforms on local sections of the time trace, covering cycles $n_{\mathrm{sweep}}$ to $n_{\mathrm{sweep}}+L-1$, where $n_{\mathrm{sweep}}$ is the starting point of the section and the section length $L=36$ is used for both cases. Fig.~\ref{fig:S3} shows representative traces and decay rates of DTC peak heights for the $\Z_2$ and two-group $\Z_3$, while data for the $\Z_2$-Ising and $\Z_3$ can be found in Ref. \cite{choi2017observation} and the main text, respectively. As shown in the figure, there is a clear difference in the functional profile between short and long $T$, transitioning from a stretched to simple exponential decay. To extract the late-time decay rate $\Gamma$, the peak height data is fitted to a single exponential decay starting from $n_{\mathrm{sweep}}=15-20$, where the transient initial decay is negligible. To account for the influence of the fit starting point, we extract the fit results from $n_{\mathrm{sweep}}=15$ to $n_{\mathrm{sweep}}=20$ and associate the mean decay rate with $\Gamma$. The errors of the fits are determined by selecting the maximum of the following two error estimates: the mean individual fit error or the standard deviation of the fit results for the set of starting values. Exemplary late-time decay rates are shown in Fig.~\ref{fig:S3}(c,f) for short and long Floquet periods and the two different Hamiltonians. For short Floquet periods, we observe that DTC order is indeed robust to perturbations, manifested as a negligible $\epsilon$-dependence of $\Gamma$. However, at long Floquet periods, DTC signals are no longer stable against the perturbations, developing a quadratic behavior as a function of $\epsilon$ with a coefficient close to 1/2. Detailed analysis of this behavior is discussed in the following section.

\section{Approach to Dephasing Regime}
\begin{figure}[h]
\includegraphics[width=0.9\textwidth]{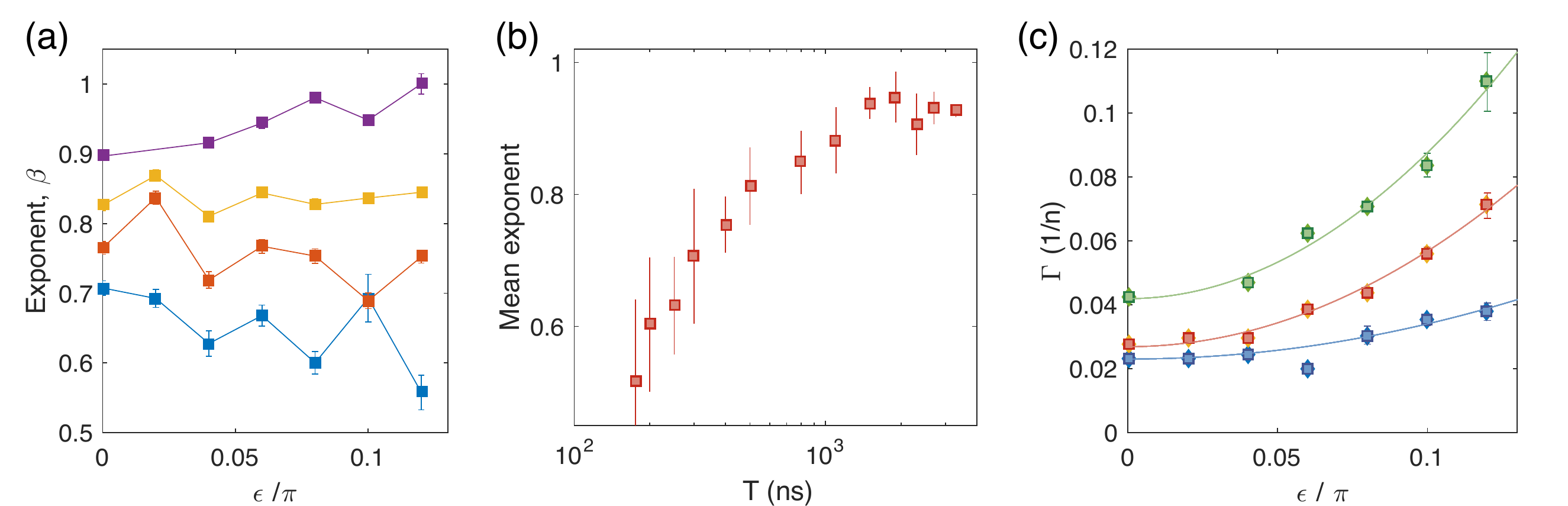}
\caption{\textbf{Crossover in the dynamics of DTC order approaching the dephasing regime.} (a) Exponent as a function of perturbation strength $\epsilon$. The exponents are extracted from a stretched exponential decay fitted to $\Z_2$ DTC signals probed at different Floquet periods: $T$ = 250 ns (blue), 400 ns (red), 800 ns (yellow), and 1900 ns (purple). (b) Mean exponent $\overline{\beta}$ as a function of $T$. (c) Late-time decay rate of $\Z_2$ DTC order measured at $T$ = 200 ns (blue), 800 ns (red), and 1900 ns (green). Solid lines are phenomenological quadratic fits.} 
\label{fig:S4}
\end{figure}

As seen in Fig.~3(a,b) in the main text, the functional profile of the DTC order decay differs depending on the length of the Floquet period $T$; at short $T$, the decay profile follows a stretched exponential, while at long $T$, it turns into a single exponential. To quantify these qualitative differences, we phenomenologically fit the DTC order decay profile using a stretched exponential function $A\exp[-(n_{\mathrm{sweep}}/n_\tau)^{\beta}]$, with exponent $\beta$ and characteristic decay constant $n_\tau$. In Fig.~\ref{fig:S4}(a), the extracted $\beta$ is displayed as a function of $T$ and $\epsilon$ for the $\Z_2$ case. As seen in the figure, $\beta$ has a more pronounced dependence on $T$ compared to its dependence on $\epsilon$. The same qualitative behavior is also observed for other cases including $\Z_2$-Ising and $\Z_3$. Therefore, we proceed to monitor the mean $\overline{\beta}$ and its errorbar at each $T$ by estimating the mean and statistical fluctuation of the local $\beta$ values extracted at different perturbations $\epsilon$. As presented in Fig.~\ref{fig:S4}(b), as $T$ increases, $\overline{\beta}$ also increases continuously to $\sim$1 consistently for all four DTC cases (see Fig.~4(b) in the main text). We attribute the saturation exponent slightly less than 1 to the convolution of the decay profile with the bare $T_1$ decay profile, the latter following a stretched exponential profile with exponent $1/2$ \cite{choi2017depolarization}. 

In addition to the changes in the functional profile, we also investigate the change in DTC stability at different Floquet periods by examining the late-time decay rate $\Gamma$ of DTC order. Fig.~\ref{fig:S4}(c) shows the DTC decay rate profiles as a function of $\epsilon$, measured at short, intermediate, and long $T$ for the $\Z_2$ case. As expected, at short $T$, the $\Z_2$ DTC order shows a robust $\Gamma$ fairly independent of perturbation strength $\epsilon$. At intermediate $T$, however, the plateau region manifesting the rigidity shrinks, and accordingly, DTC order starts to die out at a smaller perturbation than that of shorter $T$. More interestingly, at long $T$, we find that the behavior is well captured by a dephasing model predicting $\Gamma(\epsilon)=\epsilon^2/2$ up to a finite global offset $\Gamma_0$. In the main text, we present the $\Z_3$ data exhibiting similar behaviors (see Fig. 3(c) and Fig. 4(b)). 

\begin{figure}[h]
\includegraphics[width=0.75\textwidth]{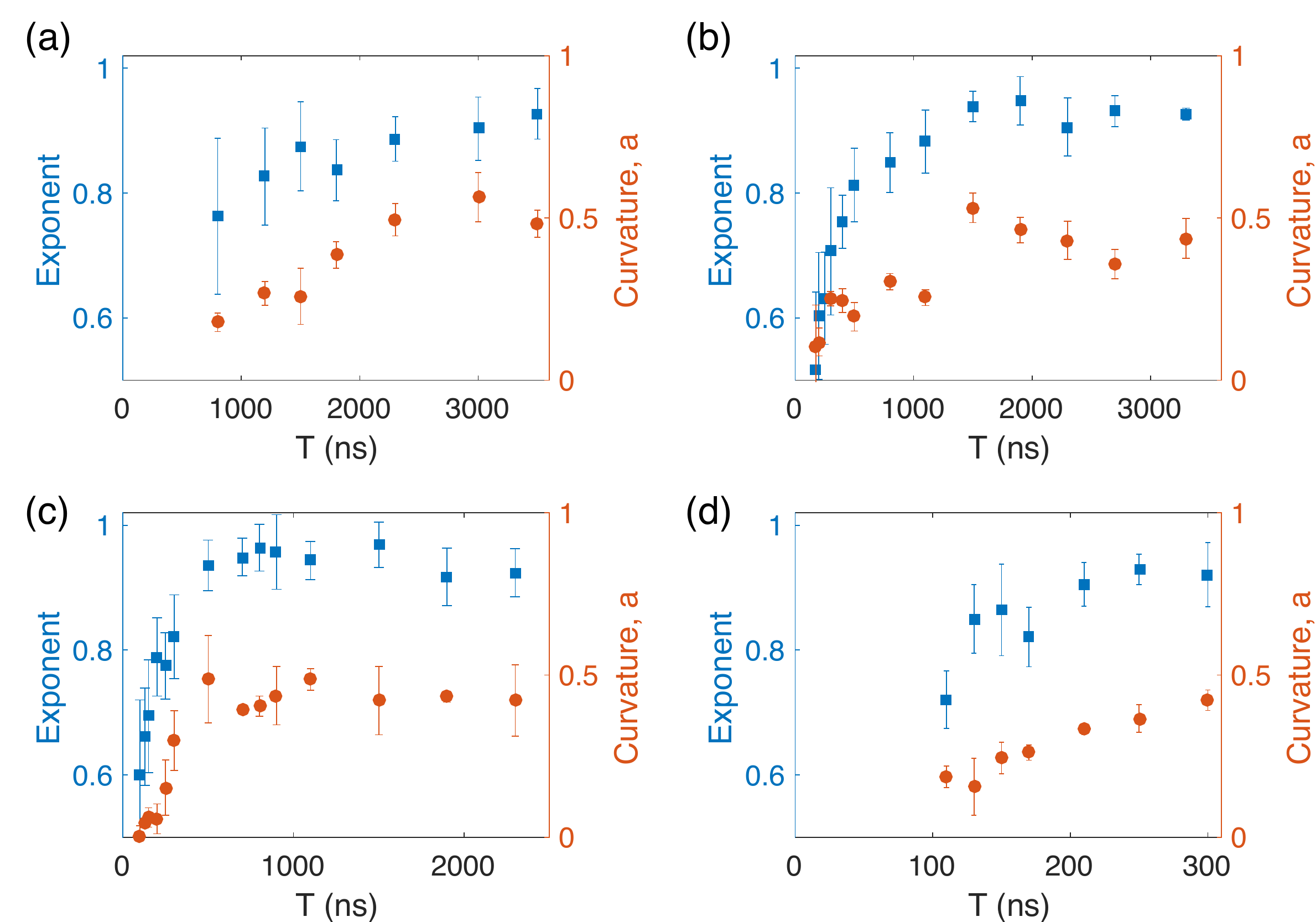}
\caption{\textbf{Correlation between late-time decay rate and functional profile of DTC order decay.} Curvature $a$ from the fit $\Gamma =  \Gamma_0 + a\epsilon^2$, and exponent $\overline{\beta}$, are compared as a function of $T$ for the different cases of (a) $\Z_2$-Ising, (b) $\Z_2$, (c) $\Z_3$, and (d) two-group $\Z_3$. We see that the curvature and exponent approach saturation at comparable interaction timescales.} 
\label{fig:S5}
\end{figure}

To identify such crossovers in the dynamics of DTC order, we perform a phenomenological fit using $\Gamma(\epsilon)=\Gamma_0+a\epsilon^2$, with a $T$-dependent offset $\Gamma_0$ and curvature $a$. The extracted $a$ values as a function of $T$ are presented in Fig.~\ref{fig:S5}, together with the exponent $\overline{\beta}$  evaluated independently from the stretched exponential fit. Surprisingly, we find a similar saturation behavior in the curvature $a$ probed as a function of $T$; the curvature $a$ also saturates around 0.5 for all cases, consistent with the simple dephasing model explaining the DTC dynamics in the limit of long $T$ (see following section). The correlation between the curvature $a$ and exponent $\overline{\beta}$ confirms that there exists a gradual crossover in the late-time DTC dynamics, approaching the dephasing regime associated with thermalization. In order to demarcate the dephasing regime in each of the different DTC realizations, we identify a transition point $T^{\star}$ where $\overline{\beta}$ increases above 0.9. The error on $T^{\star}$ corresponds to the statistical error of a saturation fit, $\beta = 1/(1 + (c_1/T)^{c_2})$, where $c_{1,2}$ are free parameters. 

\begin{figure}[h]
\includegraphics[width=1\textwidth]{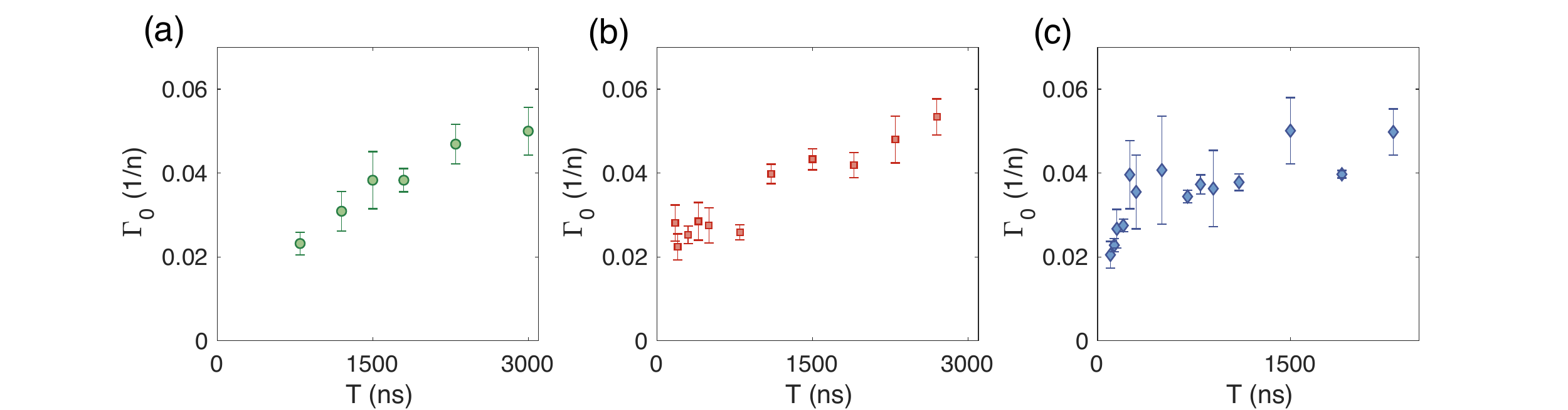}
\caption{\textbf{Dependence of offset decay rate $\Gamma_0$ on Floquet period.} (a) $\Z_2$-Ising, (b) $\Z_2$, (b) $\Z_3$ cases. } 
\label{fig:S6}
\end{figure}

As seen in Figure~\ref{fig:S6}, $\Gamma_0$ increases linearly with $T$, with a slope that is consistent with the depolarization rate $1/T_1$ of our spin system: for the $\Z_2$-Ising, $\Z_2$, $\Z_3$ cases, the inverse slopes in units of microseconds are 60(8), 86(10), and 110(26) $\mu$s, respectively. The finite offset as one extrapolates the linear curve to $T=0$ is likely a result of the finite duration of rotation pulses: for the $\Z_2$-Ising, $\Z_2$, $\Z_3$ cases, the finite offsets in units of inverse Floquet cycles are 0.01(5), 0.022(2), and 0.026(2), respectively.

\section{Universality in Long Interaction Time Regime}
We now consider the DTC response in the long interaction time regime (or equivalently, slow Floquet drive regime), where the Floquet drive energy scale $\omega_0 = 2\pi/T$ is smaller than both the on-site disorder strength and typical interaction strengths in the system. In this regime, we expect the system to be able to absorb/emit energy from/to the drive to compensate energy differences, leading to a reduced disorder strength that is smaller than typical interaction strengths in the system. Intuitively, many different spin configurations will become resonant with each other due to the significantly reduced disorder, giving rise to faster thermalization.

Phenomenologically, we may describe this limit by a simple model, in which we assume that the system effectively thermalizes during one Floquet cycle. As the dipolar interaction conserves total polarization, we model the dynamics during the interaction period as an effective dephasing, in which the spin coherence is lost and only population information remains.

First, we consider the $\Z_2$ case. In the fully-dephased limit, the microscopic form of the interaction Hamiltonian ceases to matter as long as it conserves total polarization. The action of the DTC sequence on the ensemble-averaged density matrix $\rho$ can be viewed as the alternation of a unitary rotation
\begin{align}
U_{\Z_2,\epsilon}=\begin{pmatrix}
-\sin(\frac{\epsilon}{2}) & -i\cos(\frac{\epsilon}{2})\\
-i\cos(\frac{\epsilon}{2}) & -\sin(\frac{\epsilon}{2})\\
\end{pmatrix},
\end{align}
and a dephasing channel
\begin{align}
\rho \mapsto \sum_{\mu\in\{0,1\} }|\mu\rangle\langle\mu|\rho|\mu\rangle\langle\mu|,
\end{align}
where we denote the two spin states as $\ket{0}$ and $\ket{1}$. As the dephasing channel eliminates coherences in $\rho$, we can model the time evolution in terms of population dynamics with a rate equation $p_\mu=R_{\Z_2,\mu\nu}p_\nu$, where $p_\mu=\rho_{\mu\mu}$ is the population in the spin state $\ket{\mu}$ and the evolution matrix $R_{\Z_2,\mu\nu}=|(U_{\Z_2,\epsilon})_{\mu\nu}|^2$. For the $\Z_2$ case, the long-time decay rate $\gamma_{\Z_2}$ is given by the smallest non-zero eigenvalue $\lambda_{\Z_2}$ of $R_{\Z_2}^2$ via $\exp(-2\gamma_{\Z_2})=\lambda_{\Z_2}$. Plugging in the preceding expressions for $R_{\Z_2}$ and expanding to leading order in $\epsilon$, we find that
\begin{align}
R_{\Z_2}^2=\begin{pmatrix}
\sin(\frac{\epsilon}{2})^2 & \cos(\frac{\epsilon}{2})^2\\
\cos(\frac{\epsilon}{2})^2 & \sin(\frac{\epsilon}{2})^2\\
\end{pmatrix}^2
=
\begin{pmatrix}
1-\frac{\epsilon^2}{2} & \frac{\epsilon^2}{2}\\
\frac{\epsilon^2}{2} & 1-\frac{\epsilon^2}{2}\\
\end{pmatrix}+O\left(\epsilon^3\right),
\end{align}
which has eigenvalues $1$ and $1-\epsilon^2$. The eigenvalue corresponding to an eigenvector with nonzero $S^z$ expectation value is $1-\epsilon^2\approx \exp(-2\gamma_{\Z_2})$. Hence, the decay rate in this limit turns out to be $\gamma_{\Z_2}=\frac{\epsilon^2}{2}$. This result is in good agreement with the dependence of the observed decay rate on $\epsilon$ at long interaction times.

We now consider the $\Z_3$ case, again in the limit where the system is expected to thermalize within one Floquet cycle. We use $\rho$ to denote the ensemble-averaged density matrix for the spin-1 particles. We may write the rotation matrices as
\begin{align}
U_{\Z_3,\epsilon}=\begin{pmatrix}
-\sin(\frac{\epsilon}{2}) & -i\cos(\frac{\epsilon}{2}) & 0\\
-i\cos(\frac{\epsilon}{2})  & -\sin(\frac{\epsilon}{2}) & 0\\
0 & 0 & 1\\
\end{pmatrix}
\begin{pmatrix}
1 & 0 & 0\\
0 & -\sin(\frac{\epsilon}{2}) & -i\cos(\frac{\epsilon}{2})\\
0 & -i\cos(\frac{\epsilon}{2}) & -\sin(\frac{\epsilon}{2})
\end{pmatrix}
=\begin{pmatrix}
-\sin(\frac{\epsilon}{2}) & \frac{1}{2}i\sin(\epsilon) & -\cos^2(\frac{\epsilon}{2})\\
-i\cos(\frac{\epsilon}{2}) & \sin^2(\frac{\epsilon}{2}) & \frac{1}{2}i\sin(\epsilon)\\
0 & -i\cos(\frac{\epsilon}{2}) & -\sin(\frac{\epsilon}{2})\\
\end{pmatrix}
\end{align}
and the dephasing channel as
\begin{align}
\rho \mapsto \sum_{\mu\in\{-1,0,1\} }|\mu\rangle\langle\mu|\rho|\mu\rangle\langle\mu|,
\end{align}
where the sum runs over all three spin states. The rate equation for populations has the rate-mapping matrix $R_{\Z_3,\mu\nu}=|(U_{\Z_3,\epsilon})_{\mu\nu}|^2$, which in the limit of small $\epsilon\ll\pi$, gives
\begin{align}
R_{\Z_3}^2=\begin{pmatrix}
1-\epsilon^2 & \frac{\epsilon^2}{2} & \frac{\epsilon^2}{2}\\
\frac{\epsilon^2}{2} & 1-\epsilon^2 & \frac{\epsilon^2}{2}\\
\frac{\epsilon^2}{2} & \frac{\epsilon^2}{2} & 1-\epsilon^2\\
\end{pmatrix}+O\left(\epsilon^4\right),
\end{align}
with eigenvalues $1$ and $1-\frac{3}{2}\epsilon^2$ (degeneracy 2). The resulting decay rate satisfies $\exp(-3\gamma_{\Z_3})\approx 1-\frac{3}{2}\epsilon^2$, so that $\gamma_{\Z_3}\approx \frac{\epsilon^2}{2}$. Thus, for the $\Z_3$ case, we also expect an asymptotic decay rate scaling as $\epsilon^2/2$ in the thermalizing regime.

In conclusion, we have demonstrated that, when the Floquet period $T$ is sufficiently long such that the system dynamics effectively behaves as thermalizing within each cycle, the decay rate of the subharmonic signal should scale as $\Gamma = \epsilon^2/2$. In reality, however, there will be additional decays $\Gamma_0$ due to interaction-induced dephasing associated with a finite pulse width \cite{rovny2018observation} as well as other imperfections, as discussed in the previous section.

\section{Simulations for Probing Thermalization in Long Interaction Time Regime}

To lend support to the dephasing picture in the long interaction time regime, we carry out numerical simulations based on the exact diagonalization of a many-body Hamiltonian subject to a periodic drive. More specifically, we consider the following toy model consisting of an infinite-range interacting spin-1/2 system, which captures how spin-spin interactions lead to rapid dephasing of individual spins, and ultimately thermalize the system,
\begin{align}
H(t) = \sum_i \Omega_y (t) S_i^y + \sum_{ij} \frac{J_{ij}}{\sqrt{N}} \left[\alpha (S_i^x S_j^x + S_i^y S_j^y)-S_i^z S_j^z \right],
\end{align}
where $\Omega_y(t) = (\pi + \epsilon) \delta(t-T)$ characterizes a periodic imperfect rotation of spins, $J_{ij}$ is a random coupling strength sampled from a uniform distribution, i.e., $J_{ij} \sim \mathcal{U}[-1, 1]$, and $\alpha$ is a coefficient tuning the strength of spin-exchange interactions relative to that of Ising interaction. In the following simulations, we consider only two cases of $\alpha = 1$ (both spin-exchange and Ising) and $\alpha = 0$ (pure Ising) for relevance to the experiment.

\begin{figure}[h]
\includegraphics[width=0.9\textwidth]{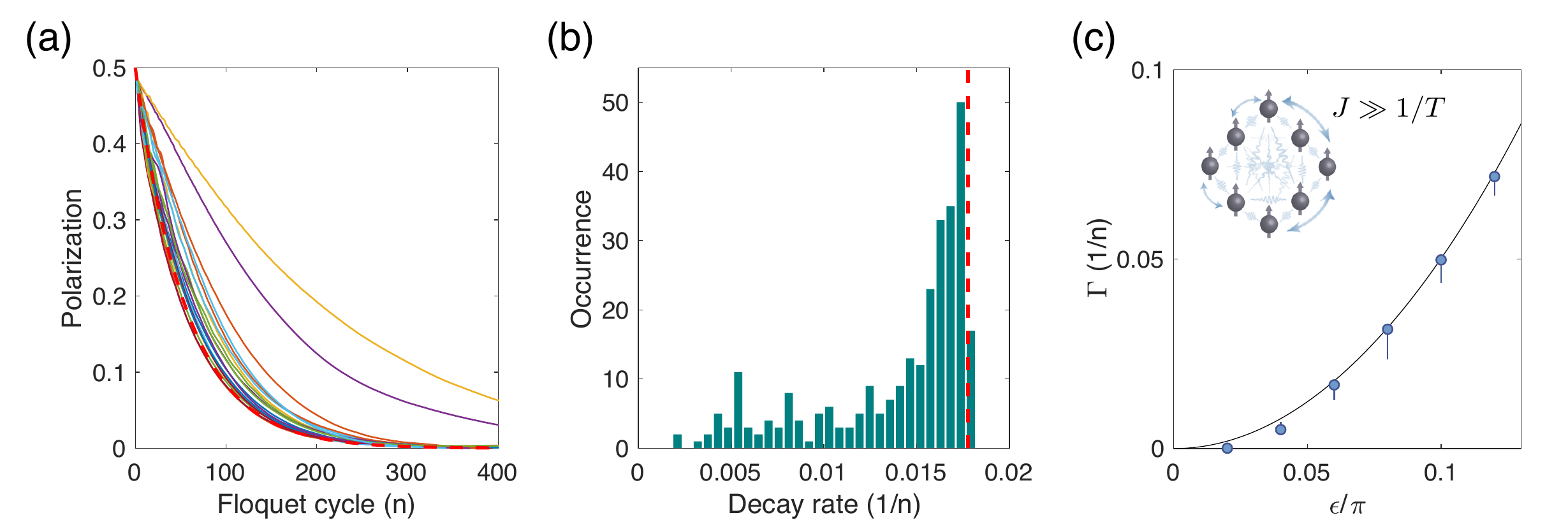}
\caption{\textbf{Simulation results for an infinite-range interacting spin system subject to a periodic drive.} (a) Twenty individual profiles of global spin polarization from different disorder realizations, probed at every even period. The red dashed line represents a simple exponential decay with a decay constant of $\Gamma = \epsilon^2/2$. (b) Histogram of late-time decay rates from the individual time traces. The vertical red dashed line indicates a position of $\Gamma = \epsilon^2/2$. (c) Most probable late-time decay rate as a function of perturbation strength $\epsilon$, extracted from (b). Errorbars indicate the standard deviation of the distribution in the histogram fitted to an asymmetric Gaussian function. In (a-c), the perturbation strength is set to $\epsilon/\pi = 0.06$. We chose a system size $N = 18$, long Floquet drive period $T = 10/J$, and included both spin-exchange and Ising interactions between the spins with $\alpha = 1$. The spins are initially polarized along the same direction ($z$-axis). } 
\label{fig:S7}
\end{figure}

Figure~\ref{fig:S7} shows the simulation results for the periodically-driven, infinite-range coupled spin system ($N = 18,~ \epsilon/\pi = 0.06$,~$JT = 10$, where $J \equiv \max{J_{ij}/\sqrt{N}}$). All spins are initially polarized along the same direction ($z$-axis) and interact with one another via both Ising and spin-exchange interactions with $\alpha = 1$. We simulate $\sim$300 disorder realizations and all of them exhibit period-doubled oscillations decaying over time with differing decay constants [Fig.~\ref{fig:S7}(a)]. For analysis, we extract the individual decay rates by fitting the late-time data ($n >40$) to a simple exponential. Interestingly, we find that a majority of the realizations display a similar decay rate very close to $\Gamma = \epsilon^2/2$ as seen in Fig.~\ref{fig:S7}(b). Repeating the same simulations with varying perturbation strength $\epsilon$, we identify the most probable decay rates as well as the standard deviation of the distribution from each histogram, and plot them as a function of $\epsilon$ [Fig.~\ref{fig:S7}(c)]. Indeed, the extracted decay rates from the simulations agree well with the expected scaling of $\Gamma = \epsilon^2/2$, consistent with the experimental observations [see Fig.~4(b) in the main text]. These numerical results substantiate our picture of thermalization in the long interaction time regime, that the dynamics can be effectively described by single-spin dephasing induced by the rest of the system acting as its own Markovian bath.  

\begin{figure}[h]
\includegraphics[width=0.5\textwidth]{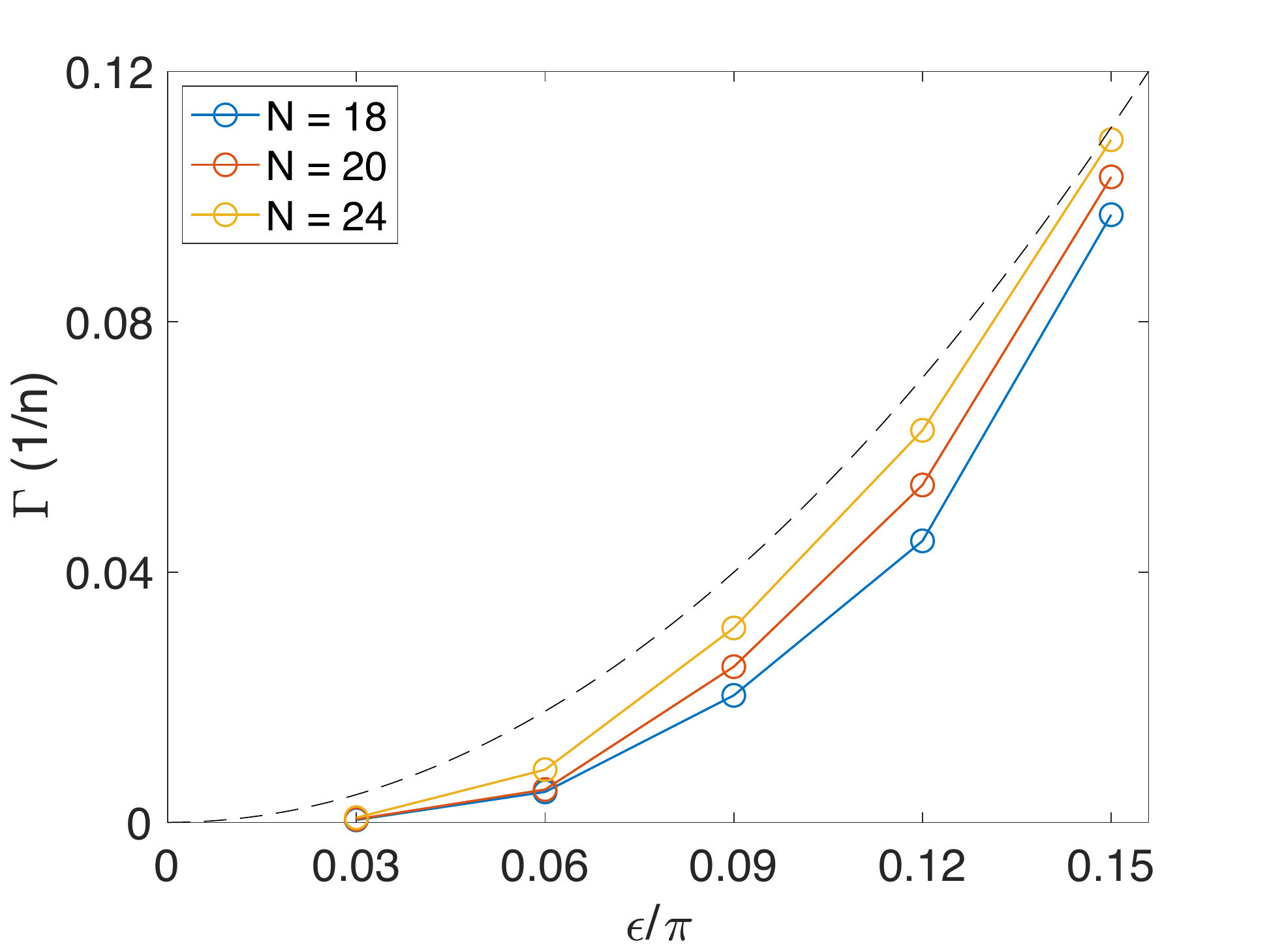}
\caption{\textbf{Simulation results for an infinite-range Ising-interacting spin system subject to a periodic drive.} (a) local spin polarization probed at every even period, $P(t) = \bra{\psi_1(t)} S_1^z \ket{\psi_1(t)}$, with an initial state $\ket{\Psi} = \ket{\psi_1} \otimes \ket{\psi_{2,\dots,N}}$, where $\ket{\psi_1} = \ket{\uparrow}$ and $\ket{\psi_{2,...,N}}$ is a $2^{(N-1)}$-dimensional complex random vector. The late-time decay rate scaling for different system sizes $N = 18, 20, 24$ are presented. The dashed line represents a dephasing fit $\Gamma = \epsilon^2/2$. We chose a long Floquet drive period $T = 10/J$, and allowed only Ising interactions between the spins with $\alpha = 0$.}
\label{fig:S8}
\end{figure}

In the case of purely Ising interactions ($\alpha = 0$), we employ a different initial state $\ket{\Psi} = \ket{\psi_1} \otimes \ket{\psi_{2,\dots,N}}$, where $\ket{\psi_1} = \ket{\uparrow}$ and $\ket{\psi_{2,...,N}}$ is a $2^{(N-1)}$-dimensional complex random vector representing a highly entangled state for the remaining $(N-1)$ spins, to be less sensitive to different disorder realizations. After each Floquet cycle, we probe the local polarization of the initially polarized spin $P(t) = \bra{\psi_1(t)} S_1^z \ket{\psi_1(t)}$. As shown in Fig.~\ref{fig:S8}, we find that the resulting late-time decay rates of subharmonic oscillations is also approaching the expected $\Gamma = \epsilon^2/2$ scaling as the system size $N$ increases. However, the finite-size-scaling approach to the Markovian regime is apparently slower than the case of spin-exchange interactions ($\alpha = 1$), as the largest system of size $N = 24$ still yields decay rates that are slower than the Markovian dephasing limit $\Gamma = \epsilon^2/2$. These observations suggest potential differences in the thermalization dynamics of systems with different types of interaction, requiring further investigations.

\newpage
\bibliography{3TDTC_supp_bibtex}